\documentstyle[multicol,prd,aps]{revtex}

\input psfig.sty

\newcommand\w{\omega}

\newcommand\D{\Delta}
\newcommand\T{\theta}
\newcommand\A{\alpha}
\newcommand\TT{\theta\theta}

\begin{document}

\title{Suspensions Thermal Noise in the LIGO Gravitational Wave Detector}

\author{Gabriela Gonz{\'a}lez}

\address{
Center for Gravitational Physics and Geometry, Department of
Physics,\\
The Pennsylvania State University, 
104 Davey Lab, University Park, PA 16802.}

\maketitle
\begin{abstract}
We present a calculation of the maximum sensitivity achievable by the
LIGO Gravitational wave detector in construction, due to limiting
thermal noise of its suspensions. We present a method to calculate
thermal noise that allows the prediction of the suspension thermal
noise in all its 6 degrees of freedom, from the energy dissipation due
to the elasticity of the suspension wires. We show how this approach
encompasses and explains previous ways to approximate the thermal
noise limit in gravitational waver detectors. We show how this
approach can be extended to more complicated suspensions to be used in
future LIGO detectors.

\end{abstract}

\pacs{04.80.Nn, 05.40.-a, 95.55.Ym}

\begin{multicols}{2}

\section{Introduction}\label{sec:intro}

Thermal noise is a fundamental limit to the sensitivity of
gravitational wave detectors, such as the ones being built in the use
by the LIGO project\cite{detectors}. Thermal noise is associated with
sources of energy dissipation \cite{PRSThNs}, following the
Fluctuation-Dissipation Theorem. Thermal noise comes in at least two
important kinds: one due to the brownian motion of the mirrors,
associated with the losses in the mirrors' material; and another due
to the suspension of the mirrors, due to the losses in the wires'
material. The limits following from these assumptions (losses due to
elastic properties of materials) are a lower limit to the noise in the
detector, since there may always be other sources of energy
dissipation in imperfect clamps, mirror attachments, etc. But the
correct calculation of the thermal noise limit is essential to the
design of detectors and diagnostics of the already-built detectors. We
will deal in this article with thermal noise of suspensions (not of
internal modes of the mirrors themselves), and assume only losses due
to the elasticity of the suspension wires.

The calculation of thermal noise can be done in several ways
\cite{JOSA},\cite{GEOThNS},\cite{Yuri},\cite{AaronPend},\cite{Matt}. All
of these follow the Fluctuation-Dissipation Theorem (FDT), but a
complication arises because in suspensions there are two sources of
energy (gravitational and elastic), but only one of them is ``lossy''
(elastic energy). Moreover, the losses in the suspension wires are
associated with their bending, and seems to be localized at
the top and bottom of the wires. The ways to include these features
into the thermal noise calculations are different enough that they
have led to some confusion among the gravitational wave
community. Also, attention has been paid mostly to the horizontal
motion of the suspension, although all modes (angular, transverse, and
vertical) appear to some degree into the detector's noise. We present
a method to calculate thermal noise that allows the prediction of the
suspension thermal noise in all its 6 degrees of freedom, from the
energy dissipation due to the elasticity of the suspension wires. We
also show how the contributions of thermal noise in different
directions can be sensed by the interferometer through the laser beam
position and direction. The results will follow from the consideration
of the coupled equations of the suspension and the continuous wire,
first presented in \cite{JOSA} for just the horizontal degree of
freedom. We show how this approach encompasses and explains previous
ways to approximate the thermal noise limit in gravitational wave
detectors. We show how this approach can be extended to more
complicated suspensions to be used in future LIGO detectors. To our
knowledge, this is the first time the thermal noise of angular degrees
of freedom is presented, and that all suspension degrees of freedom
are calculated in an unified approach.

Since the full treatment of the problem is somewhat involved, we
present first the problem without considering the elasticity of the
wire, but adding a second, lossy, energy source to the gravitational
energy in the treatment of the mechanical pendulum, and introduce the
concepts of ``dilution factors'', and ``effective'' quality
factors. We also start with one and two-degrees of freedom suspensions
instead of 6-dof. With these tools, most of the issues can be clearly
presented and then we follow to the full treatment of the LIGO
suspensions, presenting the implications for LIGO.

\section{Simple pendulum cases: dilution factors, coupled modes, 
effective quality factors.}

The full treatment of this case, considering the elastic coupling of
the wire to the suspension, was presented in \cite {JOSA}. Here, we
will present the simpler ``mechanical'' treatment of this case, which
will introduce the concepts of ``dilution factors'', and measured
vs. effective quality factors.

\subsection{A simple oscillator with a dissipative energy source}

We first recapitulate the calculation of thermal noise in the simplest
case, a suspended point mass. The potential energy is $PE=(1/2)K x^2$
and $F_K=-dV/dx=-Kx$. The kinetic energy is $KE=(1/2)M\dot{x}^2$. The
admittance to an external force $F_{ext}$ is given by
$$
Y\equiv i\w \frac{x}{F_{ext}}=i\w\frac{1}{K-M\w^2}
$$

The admittance has a pole at the system eigenfrequency $w_0^2=K/M$. If
$K$ is real, the resonance has an infinite amplitude and zero width. If
the spring constant has an imaginary part representing an energy loss,
$K\rightarrow K(1+i\phi)$, then the amplitude $|Y(\w_0)|$ is finite,
and the peak has a width determined by the complex part of the
eigenfrequency $\w_0^2=(K/M)(1+i\phi)$. The width of the peak is
characterized with a quality factor $Q=1/\phi$, and it is usually
measured from the free decay time $\tau$ of the natural oscillation at
the frequency $\w_0$: $Q=\w_0\tau/2$.

The thermal noise is proportional to the real part of the admittance,
and thus to $\phi$:
$$
\Re[Y]=\frac{wK\phi}{(K-M\w^2)^2+K^2\phi^2}
$$

We are usually interested in frequencies well above $\w_0$, since the
pendulum frequency $w_0/2\pi$ in gravitational wave detectors is
usually below 1 Hz, and the detectors have their maximum sensitivity
at ~100 Hz.  At those frequencies, the thermal noise is
\begin{equation}
x^2(\w)=\frac{4k_BT_0}{\w^2}\Re[Y]\sim\frac{4k_BT_0\w_0^2\phi}{M\w^5}
\label{simpleosc}
\end{equation}

This how we see that the measured decay of the pendulum mode can be
used to predict the suspension thermal noise at gravitational wave
frequencies.  Some beautiful examples of these difficult measurements
and their use for gravitational wave detectors are presented in
\cite{GEOmeasts}, for example.

\subsection{A pendulum with two energy sources: the dilution factor.}

Next, we consider a suspended point mass, but we now assume there two
sources of energy, gravitational and elastic, each with its own spring
constant. The potential energy is then $V=V_g+V_e=(1/2)(K_g+K_e)x^2$,
and
$$
\Re[Y]=\frac{\w(K_g\phi_g+K_e\phi_e)}{((K_g+K_e)-M\w^2)^2+(K_g+K_e)^2\phi^2}
$$

If we assume that $K_g\gg K_e$, then
$$
\Re[Y]\sim\frac{\w(K_g\phi_g+K_e\phi_e)}{(K_g-M\w^2)^2+K_g^2\phi^2}
$$
and at high frequencies
$$
x^2(\w)=\frac{4k_BT_0}{\w^2}\Re[Y]\sim
\frac{4k_BT_0(K_g\phi_g+K_e\phi_e)}{M^2\w^5}
$$
 
If $\phi_g=0$ (``gravity is lossless''), or at least $K_g\phi_g\ll
K_e\phi_e$, then
$$
\frac{4k_BT_0\w_0^2(K_g/K_e)\phi_e}{M\w^5}
$$
where $\w_0^2\sim K_g/M$. We see that is the same expression as if we
had just one energy source with a complex spring constant
$K=K_g(1+i(K_e/K_g)\phi_e)$. This is why we call the factor $K_e/K_g$
the ``dilution factor'': the elastic energy is the one contributing
the loss factor to the otherwise loss-free $K_g$, but ``diluted'' by
the small factor $K_e/K_g$. The dilution factor is also equal to the
ratio of elastic energy to gravitational energy $K_e/K_g=V_e/V_g$. The
concept of a dilution factor is very useful because it is usually
easier to measure the loss factor $\phi_e$ associated with the elastic
spring constant than the quality factor of the pendulum mode. This is
because $K_e$ is usually a function of the complex Young modulus $E$,
and the imaginary part of the Young modulus is easily measurable for
most fiber materials, and can even be found in tables of material
properties. (Of course, there are subtleties to this argument, in
particular with thermoelastic or surface losses \cite{SurfLoss}, but
we are assuming the minimum material loss).

\subsection{A point mass suspended from an anelastic wire: 
calculating the dilution factor}

This case is a particular case of the one treated in \cite{JOSA}, and
here we just mention it to present the approach taken to the full
problem, and present some new relevant aspects.

We want to include the elasticity of the wire in the equations of
motion, so we treat the suspension wire as an elastic beam, and then
we have pendulum degree of freedom $x$, plus the wire's infinite
degrees of freedom $w(s)$ of transverse motion. We define a coordinate
$s$, that starts at the top of the wire $s=0$, and ends at the
attachment point to the mirror, $s=L$. Correspondingly, we will have an
eigenfrequency, associated with the pendulum mode, and an infinite
series of ``violin'' modes. The potential energy is $PE=(1/2)\int_0^L
\rho (w'(s))^2 ds$, and the kinetic energy is $KE=(1/2)\int_0^L \rho
(\dot{w}(s))^2 ds + (1/2)M\dot{x}^2$. The solutions to the wire
equation of motion, with boundary conditions $w(0)=0$ and $w(L)=x$ are
$w(s)=x\sin(ks)/\sin(kL)$, and the equation of motion for the mass $M$
subject to an external force $F$ is
$F=-M\w^2x+Tw'(L)=-M\w^2x+(T/L)x(kL/\tan(kL))$. The admittance
$$Y=i\w \frac{x}{F}=i\w\frac{T}{L}\frac{\tan kL/kL}{1-\w^2(ML/T)(\tan
kL/kL)}$$ has a pole at the pendulum frequency $w_p^2\sim T/ML$, where
$kL\ll 1$, and an infinite number of poles at the violin mode
frequencies, at frequencies $w_n\sim (T/\rho L^2)^{1/2}(1+(\rho
L/M)n^2\pi^2)$, where $kL=n\pi+\sqrt{(\rho L/M)}/(n\pi)$.

The spring ``constant'' associated with the wire and gravity's
restoring force $K=(T/L)(kL/\tan(kL))$ is in fact a function of
frequency, although it is the usual constant $T/L$ for frequencies
below the violin modes, where $kL\ll 1$.  At frequencies above the
first violin mode, the spring function is not even positive definite,
or finite. The function $K$ is real at all frequencies because we
haven't added any source of energy loss yet.  We introduce energy loss
in the system by adding the wire elastic energy to the system, and
then assuming a complex Young modulus.  The potential energy is now
$PE=(T/2)(\int_0^L w'^2(s)ds ) + (EI/2)\int_0^L w''^2(s)ds$.  The
equation of motion for the wire is
$$T\frac{d^2w(s)}{ds^2}-EI\frac{d^4w(s)}{ds^4}+\rho\w^2w(s)=0,$$ a
fourth order equation with boundary conditions $w(0)=0$, $w'(0)=0$,
and $w(L)=x$. The wire slope at the bottom, $w'(L)$, is a free
parameter (since we are assuming a point mass), and the variation of
the Lagrangian with respect to $w'(L)$ provides the fourth boundary
condition for the wire: $w''(L)=0$. We can find an exact solution for
the wire shape as a function of $x$, trigonometric functions of $ks$,
and hyperbolic functions of $k_es$, where $k,k_e$ are solutions to
$T\kappa^2-EI\kappa^4+\rho\w^2=0$ which approximate at low frequencies
the perfect string wavenumber, $k^2\sim\rho\w^2/T$ and a constant
``elastic'' wavenumber , $k_e^2\sim T/EI$ \cite{JOSA}. The distance
$\D=\sqrt{EI/T}$ is the characteristic elastic distance over which the
wire bends, especially at top and bottom clamps. In LIGO test mass
suspensions, $\D\sim$2mm, a small fraction of $L=0.45$m.  The
approximations $k^2\sim\rho\w^2/T$ and $k_e^2\sim T/EI$ are valid for
frequencies that satisfy $\w^2\ll T^2/4EI\rho$, about 12 kHz for LIGO,
so we will use them in the remainder of this article. It is also
equivalent to $k\Delta\ll 1$.

We also use an approximate solution for the wire shape, good to order
$e^{-L/\D}$($\sim 10^{-99}$ (!) for LIGO):

\begin{eqnarray}
w(s)&=&A\left(\sin(ks)-k\D\cos(ks)\right) \nonumber \\
&& + k\D\left(Ae^{-s/\D}+Be^{-(L-s)/\D}\right)	\label{shape}
\end{eqnarray}

The coefficients $A,B$ are functions of $x$ and $k$, and thus,
functions of frequency:
\begin{eqnarray*}A&=&\frac{x}{\sin(kL)-k\D\cos(kL)} \\
B&=&xk\D.
\end{eqnarray*}

In the limit $\D\rightarrow 0$, we recover the perfect wire solution,
$w(s)=x\sin(ks)/sin(kL)$. The ratio $B/A$ measures how much more (or
less) the wire bends at the bottom than at the top. The elastic
energy is well approximated by the contribution of the exponential
terms in the wire shape, at top and bottom: $PE_e=(1/2)\int
EI(w''(s))^2ds=(1/2)T\D^2\int(w''(s))^2ds\sim (1/2)
Tk^2L(A^2+B^2)$. At low frequencies where $kL\ll 1$, the ratio
$B/A\sim \D/L \ll 1$, indicating that the wire bends much more at the
top than at the bottom (recall this is a point mass).

The equation of motion for the mass when there is an external force $F$ is

\begin{eqnarray}
F&=&-M\w^2 x+Tw'(L)-EIw'''(L) \nonumber \\
&\sim& -M\w^2 x+Tw'(L) \nonumber\\
&\sim& -M\w^2x+
\left(\frac{kT}{\tan kL}\frac{1+2k\D\tan kL}{1-(k\D)/(\tan kL)}\right)x 
\label{EOMptmass}
\end{eqnarray}

The ratio of the elastic force to the gravitational force,
$EIw'''(L)/Tw'(L)$, is of order $k\Delta\ll 1$, and thus it was
dropped. If we now consider $\Delta$ complex, then the spring function
$$K=\frac{kT}{\tan kL}\frac{1+2k\D\tan kL}{1-(k\D)/(\tan kL)}$$ is
also complex, and the admittance will have a non-zero real part. At
frequencies below the violin modes where $kL\ll 1$, we have
$K\sim(T/L)(1+\D/L)$, an expression that suggests a split between a
real gravitational spring constant $K_g=T/L$ and a complex elastic
spring constant $K_e=T\D/L^2$. However, this distinction can only be
done in the approximation $\D/L\ll 1$, and low frequencies $kL\ll
1$. In general, however, we {\em cannot} strictly derive separate
gravitational and elastic spring constants from their respective
potential energy expressions: notice that the total, complex spring
constant $\sim T/(L-\D)$ was derived from the variation of the {\em
gravitational} potential energy term, which becomes complex because we
use a wire shape involving the complex distance $\D$, satisfying the
boundary conditions.

Where the approximations $\D/L\ll 1,\,\, kL\ll 1$ are valid, we can
consider the case of two separate spring constants and thus a
``dilution factor'' for the pendulum loss, $K_{e}/K_g=\D/L\sim 1/232$,
where the numerical value corresponds to LIGO parameters in Appendix
1. However, if we numerically calculate the {\em exact } pendulum mode
quality factor, we get $1/Q_p=461/\phi$: this would be the {\em
measured} Q from a decay time of the pendulum mode, if there are no
extra losses. Did we make a ``factor of 2'' mistake? In fact, this
factor of 2 has haunted some people in the community (including
myself)\cite{Jim}, but there is a simple explanation. Since the
elastic complex spring constant is proportional to $\D$, and $\D$ is
proportional to the {\em square root} of the Young modulus, then when
we make $E$ complex $E\rightarrow E(1+i\phi)$, we get $K_{e}\sim
T\D/L^2\rightarrow (T\D/L^2)(1+i\phi/2)$. That is, we get an extra
dilution factor of two between the wire loss $\phi$ and the pendulum
loss: $\phi_p=(K_g/K_{e})\Im K_{e}/\Re
K_{el}=(K_g/K_{el})\phi/2=(\D/2L)\phi\sim \phi/464$, very close to the
actual value. This teaches us that if the spring constant of the
dissipative force is not just proportional to $E$, we will get
correction factors $\partial{\log K_e}/\partial{\log E}$. 

The thermal noise below the violin modes is well approximated by the
thermal noise of a simple oscillator, as in Eqn.\ref{simpleosc}, with
natural eigenfrequency $w_0^2=g/L$ and loss $\phi_0=\D\phi/(2L)$. We
then call $\phi_0=\D\phi/(2L)$ the ``effective'' loss, in this case
equal to the pendulum loss (but we will see this is not always the
case).

We saw that the complex spring constant $T/(\D-L)$was split into
gravitational $T/L$ and elastic $T\D/L^2$ components. However both
were derived from the {\em gravitational} force $F_g=-Tw'(L)$, since
the {\em elastic} force, $F_{el}=EIw'''(L)$, was negligible. It is
because the wire {\em shape} $w(s)$ is different due to elasticity,
that the function $Tw'(L)$ is different from the pure gravitational
expression $Tx/L$. The way we split gravitational and elastic
contributions to the spring constant and then got a dilution factor,
is only valid at low frequencies. So the argument we posed in the
previous section about a dilution factor applied to the calculation in
the thermal noise in the gravitational wave band is in priciple not
applicable here, especially when taking into account that the total
force was contributed by the variation of just the {\em gravitational}
potential energy, with the elasticity in the wire shape. However,
using the wire shape without low frequency approximations, we can
numerically evaluate the integrals that make up the potential and
elastic energies (using a {\em real} $\D$), and compare the ratio
$V_{el}/V_g=\int T(w')^2 ds/\int EI (w'')^2 ds$ with the ``low
frequency'' dilution factor $\D/L$. We show the calculation of elastic
and gravitational potential energies, and their ratio, in
Fig.\ref{PtMass}. At low frequencies, the ratio is constant, and equal
to $\D/2L$: this is the dilution factor between the wire loss $\phi$
and the pendulum loss $\phi_p$, also the one to use for a
simple-oscillator approximation of the thermal noise. It is {\em not}
the ratio of the ``gravitational'' and ``elastic'' spring constants at
low frequencies, but as we explained, we had no reason to expect that,
since $V_e/V_g\neq K_e/K_g$.  At higher frequencies, the ratio
$V_{el}/V_g$ is not constant, and it gives correctly the dilution
factors for the quality factors of the violin modes. Notice that the
loss at the violin modes increases with mode number, as noted in
\cite{JOSA}: $\phi_n=(\D/L)(1+n^2\pi^2\D/L)=(\D/L)(1+\rho L \D
\w_n^2/T)$, and this anharmonic behavior is well followed by the
energy ratio.

In summary, the concept of dilution factor is strictly true only when
the total restoring {\em force} can be split into two forces, one
lossless and one dissipative, both represented with spring
constants. In the general case, if we can only split the {\em
potential energy} into two terms, one lossless and another
dissipative, then the ratio of the energies calculated as a function
of driving frequency is the exact dilution ``factor''. Moreover, this
ratio can be calculated as a function of frequency, and then we get
the different dilution factors for all the modes in the system. This
is an important lesson that also we will use more extensively in
suspensions with more coupled degrees of freedom.

\subsection{A suspended extended mass: coupled degrees of freedom 
and observed thermal noise}

We now consider an extended mass instead of a point mass, with a
single generic dissipative energy source. The pendulum motion is
described with the horizontal displacement of its center of mass $x$,
and the pitch angle of the mass, $\T$, as in the side view of a LIGO
test mass, in Fig.\ref{Pendulum}. The kinetic energy is
$KE=(1/2)(M\dot{x}^2+J\dot{\T}^2)$. Instead of a spring constant, we
have a 2x2 spring {\em matrix}. The potential energy is
\begin{eqnarray}
PE&=&\frac{T}{2}(L\A^2+h\T^2) \nonumber \\
&=&\frac{1}{2}(K_{xx}
x^2 + 2 K_{x\T}x\T + K_{\TT}\T^2)
\label{PE2DOF}
\end{eqnarray}
where $\A=(x+h\T)/L$ and $\T$ are
the normal coordinates. The point $x+h\T$ is the point where the wire
is attached to the mirror, and the angle $\A$ is the angle the wire
makes with the vertical. If we only consider gravitational forces,
$K_{xx}=T/L,\: K_{x\T}=Th/L$ and $K_{\TT}=Th(L+h)/L$. However, we will
assume that the elements of the spring matrix can be complex, and each
has its own different imaginary part.  The
eigenfrequencies are the solutions to the equation
$(K_{xx}-M\w^2)(K_{\TT}-J\w^2)-K_{x\T}^2=0$, or
\begin{eqnarray}
2MJ\w^2_\pm =&&(MK_{\TT}+JK_{xx}) \nonumber \\ 
 && \pm \left((MK_{\TT}-JK_{xx})^2+4MJK_{x\T}^2\right)^{1/2}
\label{wpm}
\end{eqnarray}
In order to calculate $x^2(\w)$ (the Brownian motion of the center of
mass), we need to calculate the admittance $x/F$ to a horizontal force
applied at the center of mass. In order to calculate $\T^2(\w)$, we
need the admittance $\T/N$ to a torque applied around the pitch
axis. If the spring constants are complex, then the admittances are
complex and we can calculate their real parts, and the thermal noise
determined by them:
$$
x^2(\w)=\frac{4k_BT_0}{\w^2}\Re\left[i\w
\frac{1-J\w^2/K_{\TT}}{K_{xx}(1-\w^2/\w_+^2)(1-\w^2/\w_-^2)}\right]
$$
$$
\T^2(\w)=\frac{4k_BT_0}{\w^2}\Re\left[i\w
\frac{1-M\w^2/K_{xx}}{K_{\TT}(1-\w^2/\w_+^2)(1-\w^2/\w_-^2)}\right]
$$
where the eigenfrequencies are now complex:
$\w_\pm^2\rightarrow\w_\pm^2(1+i\phi_\pm)$. The quality factors
measurable from the free decay of each of the eigenfrequencies are
$Q_\pm=\w_\pm\tau_\pm/2$.

At frequencies larger than any of the eigenfrequencies, we obtain
\begin{eqnarray}
x^2(\w)&\sim &\frac{4k_BT_0}{K_{xx}\w^5}
((\w_+^2\phi_+ + \w_-^2\phi_-)  - \phi_{\TT}(K_{\TT}/J)) \nonumber \\
\T^2(\w)&\sim& \frac{4k_BT_0}{K_{\TT}\w^5}
((\w_+^2\phi_+ + \w_-^2\phi_- ) - \phi_{xx}(K_{xx}/M))
\label{xtthns}
\end{eqnarray}

The system may have the two eigenfrequencies close in value
if $h(L+h)\sim J/M$ (see Fig.\ref{wpmQpm}), but for $h\ll J/ML$, we have
$\w_-^2\sim K_{\TT}/J=Th/J$ and $\w_+^2\sim K_{xx}/ML^2=T/ML$; and for
$h(L+h)\gg J/M$, $\w_+^2\sim K_{\TT}/J=Th/J$ and $\w_-^2\sim
K_{xx}/ML^2=T/ML$. In both limits, two terms cancel in the sum of loss
factors in the formulas above ($\w_-^2\phi_-\sim\phi_{\TT}K_{\TT}/J$
for small $h$, for example) and we see that

$$
x^2(\w)\approx \frac{4k_BT_0\w_p^2\phi_{xx}}{Mw^5}
\,\,\, {\rm and}\,\,\,
\T^2(\w)\approx \frac{4k_BT_0\w_\T^2\phi_{\TT}}{J\w^5}
$$

Thus, even though it is a coupled system, thermal noise in $x$ is
always associated mostly to $K_{xx}(1+i\phi_{xx})$ and the pendulum
eigenfrequency $\w_p^2=K_{xx}/M\sim T/ML$; and thermal noise in $\T$
is always associated with $K_{\TT}(1+i\phi_{\TT})$ and the pitch
eigenfrequency $\w_\T^2=K_{\TT}/J\sim Th/J$.  For both degrees of
freedom $x$ and $\T$, we obtain the thermal noise of single-dof
systems. Unfortunately, neither limit (small or large $h$ with respect
to $J/ML$) applies to the suspension parameters in LIGO test masses,
and, more importantly, even though the approximation for the
eigenfrequencies is relatively good for most values of $h$, the
approximation we used for the losses is not (Fig.\ref{wpmQpm}). The
measurable quality factors give us $\phi_\pm$, but we need $\phi_{xx}$
and $\phi_{\TT}$ to use in the thermal noise of the pendulum, and
these cannot be precisely calculated from $\phi_\pm$ unless we know
$\phi_{x\T}$, or a way to relate it to the other loss factors. We will
do this in the next section, using the elasticity of the wire.

Notice that the forces and torques we have used to calculate the
admittances $Y_x$ and $Y_\T$, will each produce {\em both}
displacement and rotation of the pendulum. This means that the thermal
noise in displacement and angle are {\bf not} uncorrelated. This can
be exploited to find a point other than the center of mass where the
laser beam in the interferometer would be sensing less displacement
thermal noise than at the center of the mirror, as was done following
a somewhat different logic in \cite{Yuri}. If we were to calculate the
thermal noise at a point a distance $d$ above the center of mass, we
then need to calculate the admittance of the velocity of that point
($i\w(x+d\T)=i\w \chi$) to a horizontal force applied at that
point. The equations of motion are
\begin{eqnarray*}F=&(K_{xx}-M\w^2)x+K_{x\T}\T\\
Fd=&K_{x \T}x+(K_{\TT}-J_y\w^2)\T
\end{eqnarray*}
and then the thermal noise  is 
\begin{eqnarray}
\chi^2(\w)&=&\frac{4k_BT_0}{\w^2}
\Re\left[i\w\frac{(x+d\T)}{F}\right]\nonumber\\
&=&\frac{4k_BT_0}{\w^2}\Re\left[Y_{xx}+d^2Y_{\TT}+
2dY_{x\T}\right] \nonumber \\
&=&x^2(\w)+d^2\T^2(\w)+2d\frac{4k_BT_0}{\w^2}\Re(Y_{x\T})
\label{chi2}
\end{eqnarray} 
where $Y_{xx}$ is the admittance of $x$ to a pure force $F$, $Y_{\TT}$
is the admittance of $\T$ to a pure torque $N$, and $Y_{x\T}$ is the
admittance of a displacement $x$ to a pure torque $N$, equal to the
admittance of $\T$ to a pure force $F$. There is an optimal distance
$d$ below the center of mass for which the thermal noise $\chi^2(\w)$
is a minimum: this distance is $d=-\Re(Y_{x\T})/\Re(Y_{\TT})$. The
resulting thermal noise is
$$\chi^2_{\rm min}(\w)=x^2(\w)- \frac{4k_BT_0}{\w^2}\frac{(\Re
Y_{x\T})^2}{\Re Y_{\TT}}$$ which is {\em less} than the thermal noise
$x^2(\w)$ observed at the center of mass. However, the expression
obtained for the distance $d$ is frequency-dependent: that means we
have to choose a frequency at which to optimize the sampling
point. 

Summarizing, we have shown that whenever there are coupled motions,
the thermal noise sensed at a point whose position depends on both
coordinates is {\em not} the sum in quadrature of the two thermal
noise ($x^2$ and $d^2\T^2(\w)$ in our case), but a combination that
depends on the ``cross-admittance''. Moreover, the thermal noise of
each degree of freedom cannot in general be calculated just from the
measured quality factors if the modes are coupled to each other
strongly enough.

\subsection{A 2-DOF pendulum suspended from a continuum wire}

We add to the previous 2-DOF pendulum a continuum wire, to be able to
add the losses due to the wire's elasticity, and calculate modal and
effective quality factors, as well as the point on the mirror at which
we can sense the minimum thermal noise. 

If we add elasticity to the problem, as in \cite{JOSA}, the potential
energy is $PE=(T/2)(\int_0^L w'^2(s)ds + h\T^2) + (EI/2)\int_0^L
w''^2(s)ds$. The boundary conditions for the wire equation are
$w(0)=0,\,\, w(L)=x+h\T$, and $w'(0)=0,\,\, w'(L)=-\T$. The equations
of motion for the pendulum are

\begin{eqnarray}
-M\w^2 x+Tw'(L)-EIw'''(L)=&F \nonumber \\
-J\w^2\T+EI(w''(L)+hw'''(L))=&N	
\label{1wirextEOM}	
\end{eqnarray}

In order to complete the equations of motion of the pendulum, we need
the shape of the wire at the bottom end. For this, we use the shape
given by the expression in Eqn.\ref{shape}, but this time the top and
bottom weights are given by
$$A=\frac{x+(h+\D)\T}{D}$$
\begin{eqnarray*}
B=\frac{1}{kD}(&&kx(\cos(kL)+k\D\sin(kL))\\
 && + \T(\sin(kL)+k(h-\D)\cos(kL)) 
\end{eqnarray*}
with $D=\sin (kL)-2k\D \cos (kL)$ and $\D=\sqrt{EI/T}$ as we used
earlier. With the shape known, we can write the equations for $x,\T$
with a spring matrix:
\begin{eqnarray*}
(K_{xx}-M\w^2)x+K_{x\T}\T=&F \\
K_{\T x}x+(K_{\T\T}-J_y\w^2)\T=&N
\end{eqnarray*}
where the spring functions are 
\begin{eqnarray}
K_{xx}&=&Tk(\cos(kL)+k\D\sin(kL))/D \nonumber \\		
K_{\T \T }&=&T(h+\D)(\sin(kL)+k(h-\D)\cos(kL))/D \nonumber \\
K_{x\T }&=&K_{\T x}=Tk(h+\D)(\cos(kL)+k\D\sin(kL))/D \label{Kxxtt}
\end{eqnarray}

\begin{figure}
\centerline{\hbox{\psfig{figure=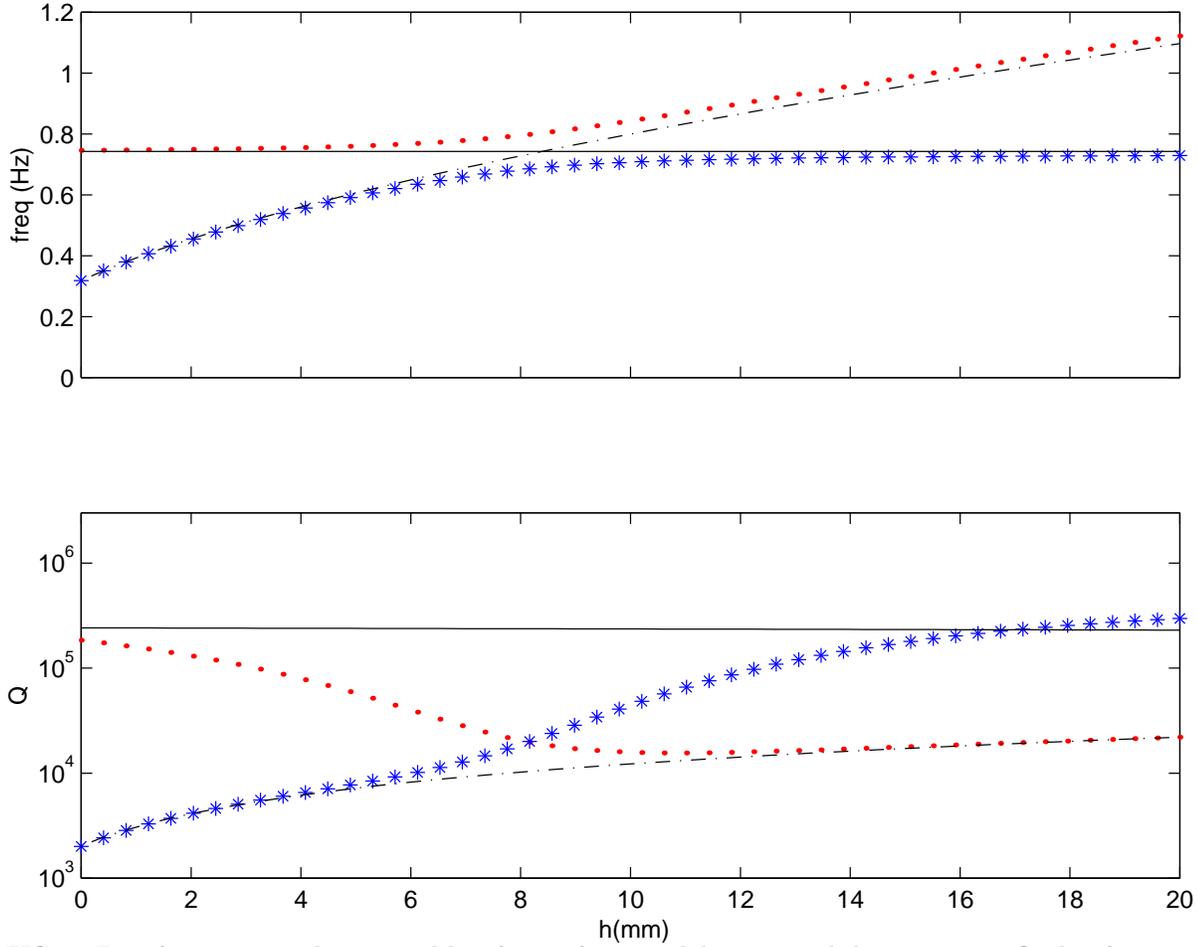,height=5in}}}
\caption{Eigenfrequencies and associated loss factors for a pendulum
suspended on one wire. Quality factors and corresponding
eigenfrequencies are represented with the same symbol (points or
stars). The lines in the top figure are given by expressions for the
pendulum frequency $f_p=\sqrt{g/l}/(2\pi)$ (solid) and pitch frequency
$f_\T=\sqrt{Th/J}/(2\pi)$ (dashed). The lines in the bottom figure
correspond to the effective quality factors for displacement (solid)
and for pitch (dashed). Approximate expressions for the effective
quality factors are $Q_p\sim L/\Delta\phi$ and $Q_\T\sim
2(h+\D)/\D\phi$, respectively.}
\label{wpmQpm}
\end{figure}

\begin{figure}
\centerline{\hbox{\psfig{figure=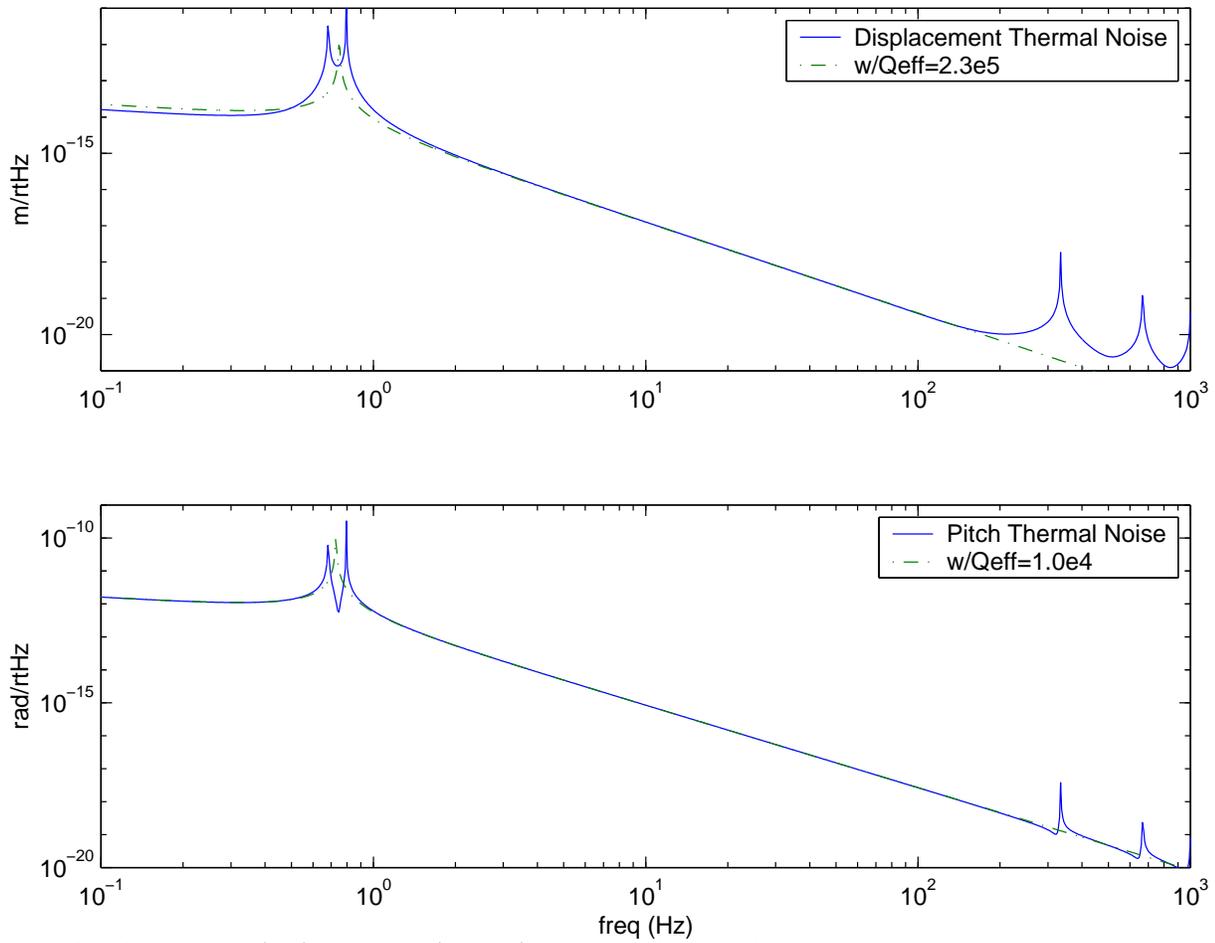,height=5in}}}
\caption{Displacement (top) and pitch (bottom) thermal noise of a
single-wire pendulum, calculated without approximations (solid line),
and approximated with a single mode with an effective Q (dotted
line). The pitch distance used was 8.3 mm}
\label{thnsasds}
\end{figure}

At this point, even though the expressions are complicated, we can
calculate the complex admittances $Y_{xx}=i\w x/F,\: Y_{\T\T}=i\w\T/N$
using a complex $E$ and $\D$.  The analytical expressions for the
admittances are quite involved, but we can always calculate
numerically the thermal noise associated with any set of
parameters. We can also calculate the widths of the peaks in the
admittance, which would correspond to measurable quality factors for
the pendulum, pitch, and violin modes. The plots presented in
Figs. \ref{wpmQpm} and \ref{thnsasds} were calculated using these
solutions.

Fig.\ref{wpmQpm} shows that the frequency and quality factor of the
pendulum and pitch modes vary significantly with the pitch distance
$h$. At frequencies close to the pendulum eigenfrequencies, the
thermal noise spectral densities show peaks at both
frequencies. However, at higher frequencies, the thermal noise
$x^2(\w)$ can always be approximated by the thermal noise of a simple
oscilaltor as in Eqn.\ref{simpleosc}, with an ``effective'' quality
factor that fits the amplitude at high frequencies to the position of
the single peak. We can similarly define an effective quality factor
for the thermal noise in $\T^2(\w)$. We show in Fig.\ref{thnsasds} the
actual and approximated thermal noise, with their corresponding
effective quality factors found to fit best at 50 Hz. The effective
quality factor at 50 Hz can be calculated as a function of pitch
distance, and we show this calculation in Fig. \ref{wpmQpm}. The
effective pitch quality factor is well approximated, for any pitch
distance, by the measurable pitch quality factor, while the pendulum
effective quality factor is close to the measurable quality factor of
the pendulum mode only at very small, or very large pitch
distances. For the LIGO pitch distance of 8mm, the measurable pendulum
quality factor is 10 times lower than the effective quality factor,
and would then give a pessimistic estimate of thermal noise amplitude.

{\em Low frequency approximation.} At low frequencies, where $kL\ll
1$, we have expressions that can help us understand how the elasticity
loss factor contributes to the effective quality factors, as well as
to the pendulum and pitch modes. We trade this gain in simplicity for
the loss of expressions valid at or above violin mode resonances.

The low frequency limit of the spring constants in Eqns. \ref{Kxxtt}
is 
\begin{eqnarray*}
K_{xx}&=&T/(L-2\D) \\
K_{\T\T}&=&T(h+\D)(L+h-\D)/(L-2\D) \\
K_{\T x}&=&K_{ x\T}=T(h+\D)/(L-2\D).
\end{eqnarray*}

 If we assume $\D$ has an imaginary part related to the material
$\phi$: $\D\rightarrow\D(1+i\phi/2)$, then we get complex spring
constants. If we use these complex spring constants in Eq.\ref{wpm} we
can calculate the loss factors of the pendulum and pitch mode,
$\phi_p$ and $\phi_\T$; and if we use them in Eqns. \ref{xtthns} we
can get the effective quality factors. Using $h/L\ll 1$ and $\D/L\ll
1$, we get $Q_{{x}{\rm eff}}=1/\phi_{xx}\sim\D\phi/L$ and $Q_{\T{\rm
eff}}=1/\phi_{\T\T}\sim\D\phi/2(h+\D)$. This represents a ``dilution
factor'' in displacement of $\D/L$ and in pitch of $\D/2(h+\D)$. These
approximations fit very well the values shown in Fig. \ref{wpmQpm}.

At low frequencies, the equations of motion for $x,\T$ can be derived
from a potential energy
$$PE_{kL\ll 1}=\frac{T}{2}\left( \frac{(x+(h+\D)\T)^2}{L-2\D} +
(h+\D)\T^2 \right).$$ Using a complex $\D$, this gives us the complex
spring constants that we can use to get mode and effective quality
factors. We would like to break up this potential energy into a
gravitational part and an elastic part, corresponding to a real
gravitational spring constant (independent of $\D$) and a lossy
elastic spring constant. We know that if we take the limit
$\D\rightarrow 0$ in $PE_{kL\ll 1}$, we obtain the regular potential
energy $PE_g$ in Eq.\ref{PE2DOF} for a 2 DOF pendulum without
elasticity. Thus, we are tempted to say that the elastic energy is the
remainder, proportional to $\D$, and thus having a complex spring
constant when we consider a complex $\D$. According to this argument,
we get $K_{gxx}=T/L$ and $K_{exx}=2T\D/L^2$. This would then give us a
dilution factor for the displacement loss $K_e/K_g\sim 2\D/L$. This is
the factor by which the imaginary part of $K_{el}$ is diluted;
however, as explained before, we pick up another factor of two due to
$K_e$ being proportional to $\sqrt{E}$. Thus, the dilution factor
between the effective quality factor and the wire quality factor is
$\D/L$. 

{\em Energy Ratios and the Dilution Factor.}We have seen that there is
another way of identifying the ``gravitational'' and ``elastic'' terms
in the potential energy, using the actual potential energy expressions
from which the equations of motion were derived: $PE_g=(T/2)(\int_0^L
w'^2(s)ds + h\T^2)$ and $PE_{el}=(EI/2)\int_0^L w''^2(s)ds$. For any
given applied force, or torque, we can solve the equations of motion
for $x,\T$ and $w(s)$. (We don't need to invoke a low frequency
approximation to do this calculation.) Then, if we calculate the ratio
of elastic to gravitational potential energy for a unit applied force,
we get a function which is frequency dependent, and is equal to the
dilution factors for the pitch mode at the pitch frequency, for the
pendulum mode at the pendulum frequency, and for the effective quality
factor at high frequencies, $\D/L$. We show the energy values and
ratios for different values of the pitch distance in
Fig. \ref{EnergyRatios}.

{\em Potential Energy Densities.}There is another interesting
calculation we can do with the solution obtained for the wire shape,
and that is to find out where in the wire the elastic potential energy
is concentrated. In other words, we want to find a relationship
between the variation of the dilution factor with frequency, and the
curvature of the wire, mostly at the top and bottom clamps. Since both
the gravitational energy and the elastic energy involve integrals over
the wire length, we can define energy densities along the wire, and
calculate a cumulative integral from top to bottom. The gravitational
potential energy has also a term $Th\T^2/2$: we define a ratio
$R=Th\T^2/(\int_0^Lw'^2 ds)$, indicating the relative contribution of
this ``pitch'' term. From Fig.\ref{CumulativeEnergies}, we observe
that the gravitational potential energy density is distributed quite
homogenously along the wire, even at the first violin mode. However, the
pitch term, which can be considered a ``bottom'' contribution,
contributes most of the gravitational energy when the system is
excited at the pitch eigenfrequency, but also in several other cases
at the pendulum frequency and at low frequencies. The elastic energy
density is concentrated at top and bottom portions of length
~$2\D$. At low frequencies, the top contributes the most; at the
pendulum eigenfrequency, the relative contributions depend strongly on $h$,
but the bottom contributes at least half the energy; at the pitch
eigenfrequency, the bottom contributes more than 99\% of the energy;
at higher frequencies, including the violin modes, top and bottom
contribute equally.

\begin{figure}
\centerline{\hbox{\psfig{figure=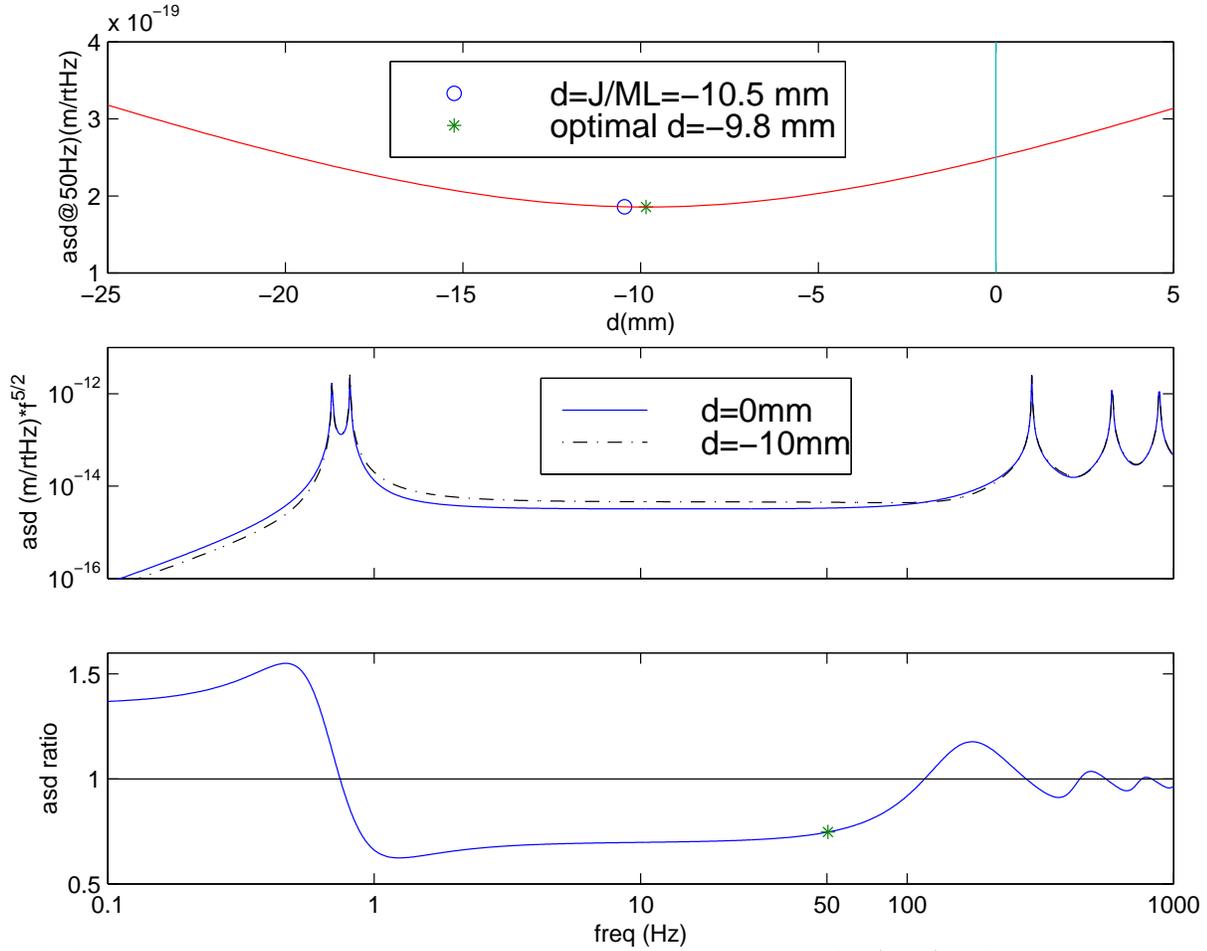,height=5in}}}
\caption{In the top figure, we plot the thermal noise amplitude
spectral density (ASD) at 50 Hz, as a function of the distance $d$ at
which the laser beam is positioned. In the middle figure, we plot the
thermal noise ASD times $f^{2/5}$ obtained for $d=0$ and for the
optimal distance for $f=50$Hz. In the bottom figure, we plot the ratio
of the two curves in the middle figure.}
\label{dmin}
\end{figure}

{\em Motion of points away from center of mass}. We discussed
previously how it was possible to find a point whose thermal noise
displacement was smaller than the thermal noise displacement of the
center of mass. Now that we have expressions for the complex spring
functions, we can find the optimal point and discuss the
differences. The cross admittance in Eq.\ref{chi2} is 
\begin{eqnarray*}
Y_{x\T}&=& i\w \frac{x}{N}= i\w \frac{\T}{F}\\
&=& -\frac{K_{x\T}}{(K_{xx}-M\w^2)(K_{\T\T}-J\w^2)}\\
&\sim& -\frac{K_{x\T}}{MJ\w^4}
\end{eqnarray*}
and then the optimal point (otimized at frequencies in the
gravitational wave band, above pendulum modes) is
$d_0=-\Re(Y_{x\T})/\Re(Y_{\TT})\sim J/ML$. Notice that even though the
optimal distance was deduced from the thermal noise expressions, which
all involve loss factors, the optimal distance only depends on
mechanical parameters. As first explained in \cite{Yuri}, the
interpretation of this distance is that when the pendulum is pushed at
that point by a horizontal force, the wire doesn't bend at the bottom
clamp, producing less losses. The fact that we can recover the result
from the FDT is another manifestation of the deep relationship between
thermal fluctuations and energy dissipation. We show in Fig\ref{dmin}
the dependence of the thermal noise at 50 Hz on the point probed by
the laser beam on the mirror, and the ratio of the thermal noise for
$d=0$ and $d=d_0$ at all frequencies. As expected, since the
integrated rms has to be the the same for any distance at which we
sense the motion, the fact that the spectral density is smaller at 50
Hz if $d=d_0$ means that the noise will be increased at some other
frequencies: this happens mainly at frequencies below the pendulum
modes.

There are many lessons to be learned from this exercise, but perhaps
the most important one is that the explicit solutions to the
equations of motion have many different important results: 

\begin{itemize}
\item using the solutions to calculate the elastic and gravitational
potential energies allows us to calculate a ``dilution function'' of
frequency, equal to the dilution factor at {\em each} of the resonant
modes of the system, as well as to the most important effective
dilution factor at frequencies in the gravitational wave band;

\item we can calculate energy densities along the wire to identify the
portions of the wire most responsible for the energy loss and thus the
thermal noise;

\item we can use low frequency approximations to find out expressions
for dilution factors that can be found using other methods, explaining 
in this way subtleties like factors of two;

\item we can calculate the admittance of an arbitrary point in the
mirror surface to the driving force, and thus find out improvements or
degradation of observed noise due to beam misalignments.

\end{itemize} 

\section{The LIGO suspensions: thermal noise of all degrees of freedom}

We will now calculate the solutions to the equations of motion for the
six degrees of freedom of a LIGO suspended test mass, and then use the
solutions to calculate the thermal nosie of all degrees of freedom, as
well as the observed tehrmal noise in the gravitational wave
detector. 

\begin{figure}
\centerline{\hbox{\psfig{figure=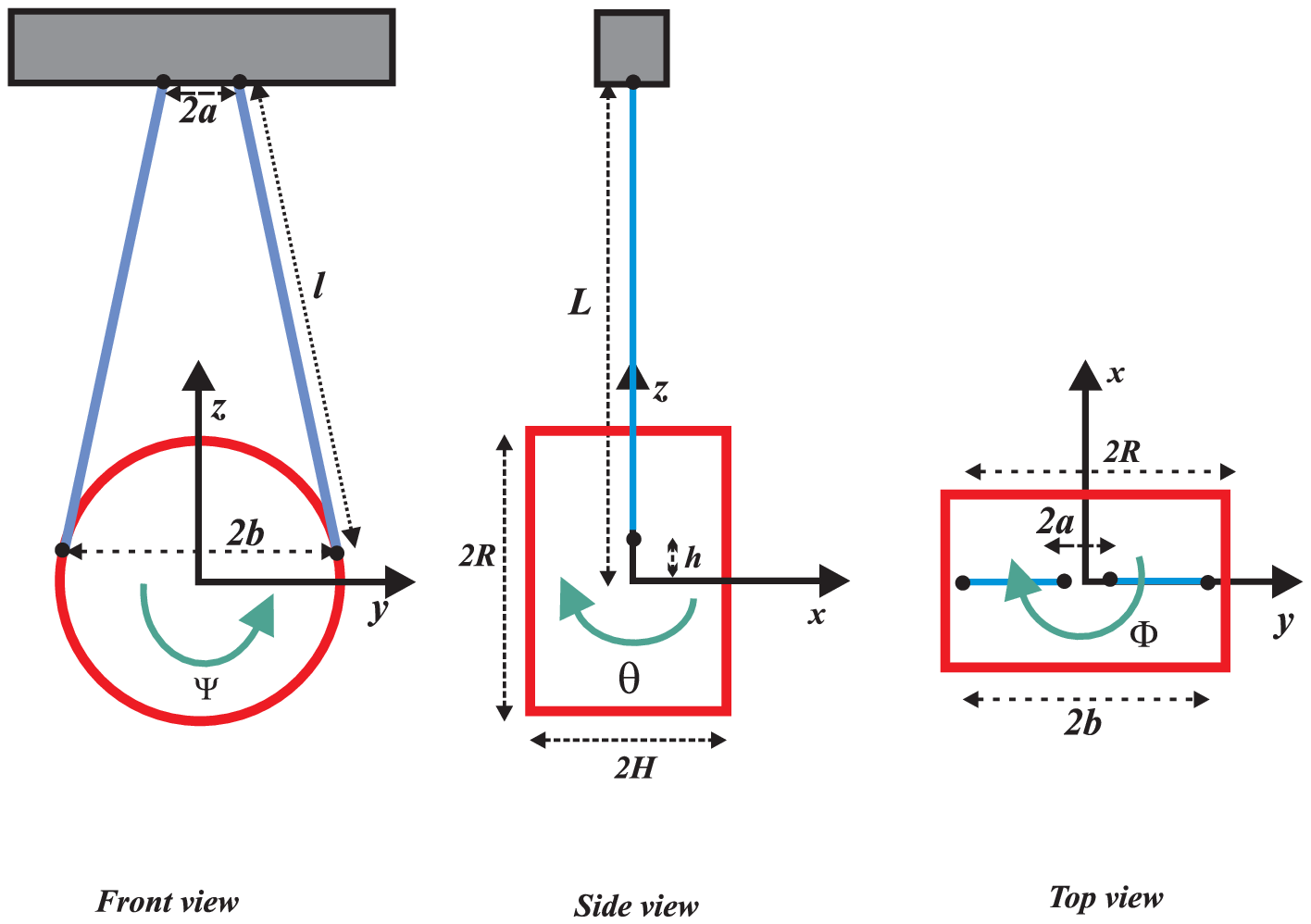,height=5in}}}
\caption{LIGO pendulum suspension, with a single wire loop, attached
slightly above the center of mass and angled towards the center.}
\label{Pendulum}
\end{figure}

The mirrors at LIGO are suspended by a single wire looping around the
cylindrical mass, attached at the top at a distance smaller than the
mirror diameter, to provide a low yaw eigenfrequency. This is
equivalent to having a mass suspended by two wires, attached slightly
above the horizontal plane where the center of mass is. The mirror's 6
degrees of freedom are the longitudinal and transverse horizontal $x$
and $y$, and the vertical $z$, displacements of the center of mass;
the pitch $\T$ and yaw $\phi$ rotations around the $y$ and $z$ axis,
respectively, and the roll $\psi$ around the longitudinal $x$ axis. We
show the coordinate system used and the relevant dimensions in
Fig.{\ref{Pendulum}. The parameters used in the calculation presented
are those for LIGO test mass suspensions (Large Optics
Suspensions). The mass of the cylindrical mirror is 10.3 Kg, the
diameter is 25cm, and the thickness 10cm. The cylindrical wires are
made of steel with density $\rho=7.8\times 10^3 {\rm kg/m^3}$ and
0.62mm diameter. We assumed a complex Young modulus $E=2.1\times 10^11
(1+10^{-3}i) {\rm kg/m^2}$. The vertical distance between the center
of mass and the top clamps is $l=45$cm, the wires are attached to the
mass a distance $h=8.2$mm above the center of mass. The distance
between the top attachment points is $2a=33.3$mm. (In the previous
examples where one wire was used, we assumed the same wire material
and the same test mirror, but we used a $0.88$mm radius, so the stress
in the wires remained constant.)

Each wire element has displacement in a 2-dimensional plane transverse
to the wire, ${\vec{w_i}}_\perp(s)$ and a longitudinal displacement
along the wire, ${w_i}_\parallel(s)$. The kinetic energy is given by
\begin{eqnarray*}
KE=\frac{1}{2}\Huge(&&\sum_{i=1,2} \int_0^L \rho |\dot{\vec{w}_i}(s)|^2 ds \\
&&+M(\dot{x}^2+\dot{y}^2+\dot{z}^2) + 
J_x \dot{\psi}^2+J_y\dot{\T}^2+J_z\dot{\phi}^2 \Huge)
\end{eqnarray*}

The potential energy is given by the sum of the axial strain energy
and the bending (transverse) strain energy in each wire:
\begin{eqnarray*}
{PE}_i=&&\frac{1}{2}\int_0^L ds 
\left(T{{w_i}'_{\perp}}^2+EA ({{w_i'}_{\parallel}})^2 \right) \\
&&+\frac{1}{2}\int_0^L ds EI\left({w_i''}_\perp\right)^2
\end{eqnarray*}
plus the energy involved in rotating the mass:
$$PE_M=
T(h\cos\alpha(\T^2+\psi^2)-b\sin\alpha(\phi^2+\psi^2)). $$

The wires will be attached at the top ($s=0$) at the coordinates
$\vec{w}_i=(0, \pm a, l)$, where $l$ is the vertical distance of the
top support from the equilibrium position of the center of mass. The
wires' transverse slopes at the top will be zero. At the bottom the
wires are clamped to the mass a distance $h$ above the center of mass,
and a distance $2b$ on the y-direction between the wires on each side
of the mass. The angle $\alpha=\arctan((b-a)/(l-h))$ is the angle at
which the wires are slanted from top to bottom when looking at the
mass along the optical axis. If $b=a$, the wires hang vertically. The
length of the wires is $L^2=(b-a)^2+(l-h)^2$. The tension in each
wire is $T=Mg\cos\alpha/2$. The position of the bottom attachments
when the mirror is moving with a motion described by
$(x,y,z,\theta,\phi,\psi)$ are $w_i(L)=(x+h\T\pm b\phi,\, y-h\psi,z\pm
b\psi)$, and the slopes at the bottom are ${w'_i}(L)=(-\T,\psi,0)$.

If we express the wire transverse and longitudinal displacements in
the $x,y,z$ coordinate system, we have
${w_i}_{\parallel}=-w_z\cos\A\pm w_y\sin\A$, and
${w_i}_\perp=\sqrt{{w_i}_x^2 + ({w_i}_y\cos\A\pm{w_i}_z\sin\A)^2}$ and
the equations of motion become non-linear.  In order to keep the
problem simple, without losing any degree of freedom, we will then
consider two different cases: (i) the wire only has displacements in
the $x$ direction, and the mirror moves in $x,\T,\phi$ degrees of
freedom; and (ii) the wire only has displacements in the $y,z$
directions, and the mirror moves in $y,z,\psi$ degrees of freedom.  We
analyze these cases separately.

\subsection{Longitudinal, Pitch and Yaw Thermal Noise}

The boundary conditions for the wires at the top are zero
displacements and slopes, and at the bottom attachment to the mass,
${w_\perp}_i(L)=x+h\T\pm\phi, {w'_\perp}_i(L)=-\T$. 

We combine the wires' transverse displacements into
$w_\pm(s)=(w_1(s)\pm w_2(s))/2$, then the boundary conditions at the
bottom are $w_+(L)=x-h\T,w'_+(L)=\T, w_-(L)=b\phi,w'_-(L)=0$. The
solutions to the wire equations of motion and the boundary conditions
are (up to order $e^{-L/\D}$) will then be $w_\pm(s)$ as in
Eqn.\ref{shape}, with
\begin{eqnarray*}
A_{-}&=&\phi b/D\\
B_{-}&=&-\phi b(\cos (kL)+k\D \sin (kL))/D\\
A_{+}&=&(x+(h+\D )\T )/D\\
B_{+}&=&-\D ( kx(\cos (kL)+k\D \sin (kL)) \\
&& + \T (\sin(kL)+k(h-\D)\cos (kL))) /D\\
\end{eqnarray*}
where $D=\sin(kL)-2k\D\cos(kL)$.  The equations for the mass
transverse dof subject to a force $F_x$ and torques $N_y,N_z$, are
\begin{eqnarray}
Fx&=&-M\omega ^{2}x-2EIw_+'''(L)+2T\theta\nonumber \\
N_y&=&-J_{y}\omega ^{2}\theta +2EI\left( w_+''(L)+hw_+'''(L)\right) 
\nonumber \\
&& +2Th\theta (1-\cos \alpha )  \nonumber \\
N_z&=&-J_{z}\omega ^{2}\phi -2EIbw_-'''(L)-2Tb\phi \sin \alpha 
\label{2wiresxtEOM}
\end{eqnarray}

We see that the combination $w_+(s)$ is associated with the $x,\T$
degrees of freedom just as for the single wire case, while the
combination $w_-(s)$ is associated with the yaw degree of freedom
$\phi$. This is easily understood when imagining the wires moving back
and forth ``in phase'' ($w_-=0$), producing displacement and pitch but
nt yaw; while if they move back and forth in opposition ($w_+=0$),
then the only effect is into mirror's yaw. Thus, we can solve the
equations for $w_-(s)$ and $\phi$ separately from the equations for
$\w_+(s),\,x$ and $\T$: we will do so in the next parapgraphs.

{\bf Yaw angular thermal noise.} The admittance of yaw $\phi$ to a
torque $N_z$ is
$$Y_\phi=i\w\frac{1}{K_\phi-2Tb\sin\A-J_z\w^2}$$ with
$K_\phi=(2Tb^2/L)(kL)(\cos(kL)+k\D\sin(kL))/(sin(kL)-2k\D\cos(kL))$. As
usual, when $\D=\sqrt{EI/T}$ is complex, the admittance is complex and
using the fluctuation-dissipation theorem we obtain
$\phi^2(f)=(4k_BT_0)\Re(Y_\phi)/\w^2$. 

The effect of the tilted wires with $\A=(b-a)/L > 0$ is to lower the
restoring force, and thus the resonance frequency: in LIGO
suspensions, the frequency is 0.48 Hz instead of 1.32 Hz if
$b=a$. However, since $\sin\A\neq 0$ decreases the gravitational
restoring force but not the elastic force, the dilution factor $\sim
K_e/K_g$ increases, and so does the thermal noise. At frequencies
where $kL\ll 1$, $K_\phi\sim (2Tb^2/L)(1+2\D/L)$. The thermal noise at
frequencies below the violin modes, where $kL\ll 1$, is well
approximated by the thermal noise of a single oscillator with
resonance frequency $w_\phi^2=2Tab/JL$ and quality factor
$Q_\phi=(L/\D)(a/b)(1/\phi_w)$, where $E\rightarrow
E(1+i\phi_w)$. Thus, the ``dilution factor'' is $(\D/L)(b/a)$ (=1/72
for LIGO parameters, where $b/a=7.5$). As in the case of a simple
pendulum, this dilution factor is half of the ratio of the elastic
spring constant $K_e=(2Tb^2/L)(2\D/L)$ to the gravitational spring
constant $K_g=2Tab/L$, because the elastic spring constant has an
``extra'' dilution factor of 2: if $E\rightarrow E(1+i\phi_w)$, then
$\D\rightarrow \D(1+i\phi_w/2)$ and $K_e\rightarrow
K_e(1+i\phi_w/2)$. However, the ratio of elastic potential energy
$\int_0^L EI (w''_\perp)^2 ds$ to gravitational potential energy
$V_g=\int_0^L T (w'_\perp)^2 ds-Tb\sin\A\,\phi^2$ is equal to the
``right'' dilution factor $(b/a)(\Delta/L)\sim 1/44$ at low
frequencies, as shown in Fig. \ref{YawDilutionFactor}. The energy
ratio also gives us the right dilution factor at the violin
frequencies ($2\D/L$).

The yaw angular thermal noise may be seen in the detectors'
gravitational wave signal if the beam hits a mirror at distance $d$ to
either side of the center of mass, or if it hits the mirror in a
direction an angle $\gamma$ away from longitudinal. Considering both
cases, the sensed thermal noise will be given by
$\chi^2(\w)=(d\cos\gamma+H\sin\gamma/2)^2\phi^2(\w)$, where $H$ is the
thickness of the mirror:
\begin{eqnarray*}\chi^2(\w)&=&(d\cos\gamma+(H/2)\sin\gamma)^2\phi^2(\w)\\
&\sim&\frac{4k_BT}{\w^5}(d\cos\gamma+(H/2)\sin\gamma)^2
\frac{2Tb^2}{J_xL}\frac{\D}{L}\phi
\end{eqnarray*}
where the approximation is valid
between the pendulum mode and the first violin mode, ~1Hz-50Hz. At
160 Hz, where the maximum sensitivity of $\sqrt{h^2(f)}=2.5\times
10^{-23}/\sqrt{\rm Hz}$ is expected, the yaw thermal noise is
$\sqrt{\phi^2}(160 {\rm Hz})=2.9\times 10^{-19} {\rm
rad/\sqrt{Hz}}$. If the yaw thermal nose is to be kept an order of
magnitude below the dominant noise source, then it is required that
$d\leq$1cm and $\gamma\leq 10^o$.

Notice that the mirror will always be aligned normal to the laser beam
to make the optical cavities resonant; however, what matters is the
beam direction with respect to the coordinate system defined by the
local vertical and the plane defined by the mirror in
equilibrium. Presumably there will be forces applied to align the
mirror, but in principle they have no effect on the response of the
mirror to an oscullatory driving force such as the one we imagine in
the beam's direction, to calculate the admittance. Thus the
requirement on $\gamma\leq 10^o$ is on the position of the mirror {\em
when there are no bias forces acting}, with respect to the ultimate
direction of the beam. The beam's direction must be within ~1$\mu$rad
of the normal to the {\em aligned} mirror to keep the beam aligned on
mirrors 4km apart, but that doesn't mean that the mirror has not been
biased by less than $10^o$ to get it to the final position. 

{\bf Pitch and displacement thermal noise.} We now solve the equations
for $w_+(s),\T$ and $x$. The equations for these degrees of freedom in
Eqns. \ref{2wiresxtEOM} are exactly the same as for the pendulum
suspended on a single wire (Eqn. \ref{1wirextEOM}), except for the
addition of a softening term to the torque equation, due to the tilted
wires; and factors of two due to the two wires (with about half the
tension) instead of a single wire. The extra term in the torque
equation is a negligible contribution to the real part of $K_{\T\T}$,
at the level of 1\% for LIGO parameters. Therefore, the conclusions we
obtained, with respect to the optimization of the beam location on the
mirror, and the difference between effective and measurable quality
factors, are equally valid here. The spring constants we obtained in
Eqns. \ref{Kxxtt} involve now a factor $2T=Mg\cos\A$, instead of
$T=Mg$ for a single wire. The elastic distance is however determined
by the tension in {\em each} wire,
$\D=\sqrt{EI/T}=\sqrt{2EI/Mg\cos\A}$. To keep the stress in the wires
constant, the cross section area of a single supporting wire is twice
the area of two each of two supporting wires. Thus, the effective
quality factor determining the thermal noise for the displacement
thermal noise $x^2(f)$ has a smaller dilution factor of $\D/L=1/326$,
instead of 1/231 for a single wire. This is the well-know effect of
reducing thermal noise by increasing the number the wires. The
dilution factor for pitch is $1/14$, considerably higher than the
dilution factor for yaw, $1/44$.

The displacement thermal noise at 160 Hz is $\sqrt{x^2(160{\rm
Hz})}=1.1\times 10^{-20} {\rm m}/\sqrt{\rm Hz}$, limiting the detector
sensitivity to $h=5.6\times 10^{-24}/\sqrt{\rm Hz}$. This is expected
to be lower than the thermal noise due to the internal modes of the
mirror mass, not considered here \cite{Aaron}. The pendulum thermal
noise could be reduced by a factor $\sqrt{2}$, or about 40\%, if the
beam spot was positioned at the optimal position on the mirror.  Since
pendulum thermal noise is not the dominant source noise, but the
detectors' shot noise would increase due to diffraction losses, it is
not advisable for LIGO to proceed this way. However, these
considerations should be taken into account for future detectors,
where thermal noise may be a severe limitation at low frequencies.

The pitch angular noise at 160 Hz, is $\sqrt{\T^2(f)}=8.9\times
10^{-19}{\rm rad}\sqrt{\rm Hz}$. Its contribution to the sensed motion
has to take into account the coupling with displacement, and we will
do this in detail inthe last section.

\subsection{Vertical, transverse displacement and roll}

We are now concerned with the mirror motion in its $y,z$ and $\psi$
degrees of freedom. The potential energy is $PE_i=(1/2)\int_0^L \left(
T({w'_i}_\perp)^2 + EI ({w''_i}_\perp)^2 +EA
({w'_i}_{\parallel})^2\right)ds$ for each wire, plus
$T(h\cos\A-b\sin\A)\psi^2$. Notice that due to the tilting of the
wires, the ``transverse'' $w_\perp$ and ``axial'' $w_\parallel$
directions are not $y$ and $z$, but rotations of these directions by
the wire tilt angle $\A$. We define ${w_i}_\perp$ for each wire
pointing ``out'' (and thus in opposite directions if $\A=0$), and
${w_i}_\parallel$ pointing down along the wire, from top to bottom.

The boundary conditions at top are
${w_i}_\perp(0)=0,{w'_i}_\perp(0)=0,{w_i}_\parallel(0)=0$, and at the
bottom, ${w_i}_\perp(L)=\pm(y\cos\A-d\psi)+ z \sin\A$,
${w'_i}_\perp(L)=\pm\psi$, and ${w_i}_\parallel(L)=\pm(y\sin\A-c\psi)-
z\cos\A$, where we defined two new distances $c=h\sin\A+b\cos\A$ and
$d=h\cos\A-b\sin\A$.  If we define as earlier, sums and differences of
the two wires shape functions, $w_\pm(s)=(w_1(s)\pm w_2(s))/2$, then
the equations of motion for the mirror degrees of freedom, when
subject to external forces $F_y,F_z$ and a torque $N_x$, are

\begin{eqnarray*}
F_y&=&-M\w^2y+2(T{w'_-}_\perp(L)-EI{w'''_-}_\perp(L))\cos\A\\
&&+2EA{w'_-}_\parallel\sin\A\\ 
F_z&=&-M\w^2z+2(T{w'_+}_\perp(L)-EI{w'''_+}_\perp(L))\sin\A\\
&&-2EA{w'_+}_\parallel\cos\A\\
N_x&=&-J_x\w^2\psi-2EI(d{w'''_-}_\perp(L)+{w''_-}_\perp(L))\\
&&+2EAc{w'_-}_\parallel(L)
\end{eqnarray*}

The solution for the wires' transverse motion ${w_\pm}_\perp(s)$
satisfying the boundary conditions up to order $e^{-L/\D}$ and $k\D$
are of the same form as in Eqn. \ref{shape}, with top and bottom
weights equal to

\begin{eqnarray}
A_-&=&(y\cos\A-(d+\D)\psi)/D \nonumber\\
B_-&=&-((y\cos\A-d\psi)(cos(kL)+k\D\sin(kL))\nonumber\\
&&-\psi(\sin(kL)-k\D\cos(kL))/k)/D \label{AB-}\nonumber\\
A_+&=&z\sin\A/D \nonumber\\
B_+&=&-z\sin\A(cos(kL)+k\D\sin(kL))/D \label{AB+}
\end{eqnarray}

The axial wire motion is 
\begin{equation}
{w_\pm}_\parallel={w_\pm}_\parallel(L)\frac{\sin(k_zs)}{\sin(k_zL)}
\label{wpar}
\end{equation}
where ${w_-}_\parallel(L)=y\sin\A-c\psi$ and
${w_+}_\parallel(L)=-z\cos\A$. The wavenumber functions are
$k^2=\rho\w^2/T$, $k_z^2=\rho\w^2/EA$.

Even though the tilting of the wires produces more complicated
formulas than in the pendulum-pitch-yaw case, the equations for the
vertical motion decouple from the equations from the transverse
pendulum displacement and roll, similar to yaw decoupling from
pendulum and pitch. As before, if the wires move in phase (transverse
or axially or both), they produce only vertical motion; but if they
move in opposition, they produce side to side motion plus rotation
around the optical axis. We analyze the two decoupled systems
separately.

{\bf Vertical thermal noise}. Once we have solved the wire shape (from
Eqns\ref{shape},\ref{AB+}, and \ref{wpar}), we can write the equation
of motion of the wire vertical displacement as
\begin{eqnarray*}
F_z=-M\w^2z&&+2\left(\frac{EA}{L}\frac{k_zL}{\tan(k_zL)}\cos^2\A \right.\\
&&\left.+Tk\frac{\cos(kL)+k\D\sin(kL)}{\sin(kL)-2k\D\cos(kL)}\sin^2\A\right)\,z
\end{eqnarray*}
or
$$F_z=-M\w^2z+(K_T\sin^2\A+K_E\cos^2\A)z$$ where we defined $K_T$ as
the spring constant that was used in the pendulum-pitch case, and
$K_E\sim 2EA/L$ the spring constant of the wire. For LIGO parameters,
and for usual wires, $K_T/K_E\ll 1$.

If the wires are not tilted and $\A=0$, we recover the simple case of
vertical modes of a mirror hanging on a single wire. The restoring
force is elastic and proportional to $E$, so there is no dilution
factor. The term added because of the wire tilting is a gravitational
restoring force, much smaller than the elastic restoring force. Since
it is also mostly real when considering a complex Young modulus $E$,
it will not change significatively the loss terms, and thus the
thermal noise. The wire tilting does add, however, the violin modes to
the vertical motion, and it slightly decreases (by a factor
$\cos\A=0.97$) the frequency of the lowest vertical mode.

The vertical thermal noise at 160 Hz is $\sqrt{z^2(f)}=3.1\times
10^{-18} {\rm m/\sqrt{Hz}}$, 260 times the pendulum thermal
noise. This is due to the lower quality factor, and the higher mode
frequency. However, vertical noise is sensed in the gravitational wave
interferometer through the angle of the laser beam and the normal to
the mirror surface, which is not less than the Earth's curvature over
4km, (0.6 mrad). At the minimum coupling (0.3 mrad for each mirror in
the 4km cavity), the contribution due to vertical thermal noise is
10\% of the pendulum thermal noise. In advanced detectors, vertical
modes are going to be at lower frequencies due to soft vertical
supports, like in the suspensions used in the GEO600 interferometer,
but the ratio of quality factors is just the mechanical dilution
factor, so the contribution of vertical thermal noise to sensed motion
will be of order $3\times 10^{-4}\times \sqrt{L/\D} \times
{f_z/f_x}\sim 3\times 10^{-4}\times (EL^2/Mg)^{1/4}$, not necessarily
a small number!

{\bf Side pendulum and roll}. The side motion and roll of the pendulum
are not expected to appear in the interferometer signal, but it is
usually the case that at least the high quality-factor resonances do
appear through imperfect optic alignment.  The equations for the
system $y,\psi$ can be written as
\begin{eqnarray*}
(K_{yy}-M\w^2)y-K_{y\psi}\psi&=&F_y\\
(K_{\psi\psi}-J_x\w^2)\psi-K_{\psi y}y&=&N_x
\end{eqnarray*}
with 
\begin{eqnarray}
K_{yy}&=&K_T\cos^2\A +K_E\sin^2\A\nonumber\\
K_{\psi\psi}&=&K_Ec^2+K_Td(d+\D)\nonumber\\
&&+2T(d+\D)\frac{\sin(kL)-k\D\cos(kL)}{\sin(kL)-2k\D\cos(kL)}\nonumber\\
K_{y\psi}=K_{\psi y}&=&K_T(d+\D)\cos\A
+K_Ec\sin\A
\label{Kyy}
\end{eqnarray}

Within the (very good) approximation $K_T/K_E\ll 1$, we can
prove that the eigenfrequencies are $w_y^2=K_T/M,f_y=0.75$ Hz and
$\w_\psi^2=K_E b^2/J,f_\psi=19$ Hz. The corresponding loss factors are
$\phi_y\sim\D\phi/2L$ and $\phi_\psi\sim\phi$.

If the wires are perfectly straight, the thermal noise of the
side-to-side pendulum motion below the violin modes $y_0^2(f)$ is well
approximated by that of a simple oscillator with eigenfrequency $\w_y$
and a dilution factor $\D/2L$. The thermal noise of the roll angular
motion $\psi^2(f)$ does not depend much on the wire tilt, and is well
approximated (below violin modes) by the thermal noise of a simple
oscillator with eigenfrequency $w_r^2=2EA/J_xL$ and quality factor
equal to the free wire's quality factor (there is no dilution factor).

For any small wire tilt, however, the spring constant $K_{yy}$ in
Eqn.\ref{Kyy} has a large contribution of $K_E$, and thus the
fluctuations increase as $y^2(f)\sim y_0^2(f)\cos^2\A+
(J_x/Mb)^2\psi^2(f)\sin^2\A$, as seen in Fig. \ref{sideroll}. Since
the roll eigenfrequency is higher than the side pendulum
eigenfrequency, and its quality factor is lower, this is a significant
increase in the thermal noise spectral density $\psi^2(f)$, about a
factor of 100 for LIGO parameters. However, the gravitationalw ave
detectors are mostly immune to the roll degree of freedom, as we will
see later.

Also, the tilting of the wires introduces violin modes harmonics of
$f_n^{(2)}=\sqrt(EA/\rho)/(2L)\sim n\times$6kHz apart from the usual
harmonics $f_n^{(1)}=\sqrt(T/\rho)/(2L)\sim n\times 300$Hz.

The violin modes that are most visible in the roll thermal noise are
the harmonics of $\sqrt(EA/\rho)/(2L)$ (and are strictly the only ones
present if the wires are not tilted).

\subsection{Violin Modes}

We have explored the relationship between quality factors of pendulum
modes and the ``effective'' quality factor needed to predict the
thermal noise of any given degree of freedom (seen only in sensitive
interferometers). For the most important longitudinal motion, we've
seen that the effective quality factor is approximately equal to $Q_w
L/\D$, where $Q_w=1/\phi_w$ is the quality factor of the free wire,
related to the imaginary part of the Young modulus. Under ideal
conditions where the pendulum mode is far away from the pitch mode in
frequency, its quality factor is close to the effective quality
factor, but as we have seen, the errors may be as large as 50\%.

The violin modes, approximately equal to $f_n=n\sqrt{T/\rho}/2L$,
appear in the horizontal motion of the pendulum in both directions,
along the optical axis and transverse to it. The violin modes show
some anharmonicity, as pointed in \cite{JOSA},with the frequencies
slightly higher than $n\sqrt{T/\rho}/2L$, and the quality factors
degrading with mode number. The complex eigenfrequencies are the
solutions to the equation
$$kL=n\pi+\arctan(2k\D/(1-(k\D)^2))$$
with $k^2=\rho\w^2/T$. This means that if we measure the
quality factors of violin modes, and they follow the predicted
anharmonic behavior in both directions, we can assume the losses are
only limited by wire losses. We can then predict the thermal noise in
the gravitational wave band with more confidence, having also
consistency checks with the quality factors measured at the pendulum
modes. The thermal amplitude of the peaks at the violin modes follows
a simple $1/f^{5/2}$ law, corrected by the change in longitudinal
Qs. We show all these features in Fig.\ref{ViolinModes}.

\subsection{Total Pendulum Thermal Noise in LIGO I}

The right way to calculate the total thermal noise observed in the
interferometer signal is to calculate the pendulum response to an
applied oscillating force in the direction of the laser beam. The
pendulum responds to a force in all its six degrees of freedom, but
the motion we are sensitive to is the motion projected on the laser
beam's direction. 

If the laser beam is horizontal, and its direction passes through the
mirror center of mass, it will be sensitive to only longitudinal
displacement and pitch motion. If the beam is not horizontal, and for
example is tilted up or down by an angle $\gamma$, (but still going
through the center of mass), we imagine an applied force $F$ applied
in the beam's direction, with a horizontal component $F\cos\gamma$ and
a vertical component $F\sin\gamma$. The motion we are interested in is
$x\cos\gamma+z\sin\gamma$, since it is the direction sensed by the
laser beam. The admittance we need to calculate is then
$Y=i\w(x\cos\gamma+z\sin\gamma)/F=
i\w((x/F_x)\cos\gamma+(z/F_z)\sin\gamma) =
i\w(Y_x\cos^2\gamma+Y_z\sin^2\gamma)$, where $Y_x$ is the response of
the horizontal displacement to a horizontal force and $Y_z$ is the
vertical response to a vertical force. A force applied in the beam's
direction, with magnitude $F_0$, will have components in all 3 axes,
and torques around all 3 axes too:
$\vec{F}=F_0(\alpha_x,\alpha_y,\alpha_z)$ and
$\vec{N}=F_0(D_\psi,D_\theta,D_\psi)$. The motion we are interested is,
in general,
$\chi=\alpha_xx+\alpha_yy+\alpha_zz+D_\theta\theta+D_\phi\phi+D_\psi\psi$. If
the motion in all 6-DOF was uncoupled, then each dof responds to just
one component of the force or the torque, and the admittance we need
would be just the sum of the admittances, each weighted by the square
of a factor $\alpha_i$ or a distance $D_i$. However, only $z$ and
$\phi$ are decoupled dof from the rest, and $x,\theta$ and $y,\psi$
form two coupled systems, for which we need to solve the response to a
forces and torques together. 

We define an admittance $Y_{x\theta}=i\w(x/N_y)$ as the admittance of
displacement $x$ to an applied torque $N_x$, and
$Y_{\theta_x}=i\w(\theta/F_x)$ as the admittance of pitch to an
applied horizontal force , and similar quantities
$Y_{y\psi},Y_{\psi,y}$. Then, the total thermal noise sensed by the
laser beam is
\begin{eqnarray*}
\chi^2(f)&=\frac{4k_BT_0}{\w^2}\Re&\left(
\A_x^2Y_{xx}+\A_xD_\T(Y_{x\T}+Y_{\T x}) + D_\T^2 Y_{\T\T} \right.\\
&&+\A_y^2Y_{yy}+\A_yD_\psi(Y_{y\psi}+Y_{\psi y}) + D_\psi^2 Y_{\psi\psi}\\
&&\left.+\A_z^2Y_{zz} + D_\phi^2 Y_{\phi\phi}\right)
\end{eqnarray*}
or
\begin{eqnarray*}
\chi^2(f)&=&\A_x^2 x^2(f)+ D_\T^2 \T^2(f)+
\A_xD_\T\frac{4k_BT_0}{\w^2}\Re(Y_{x\T}+Y_{\T x}) \\ 
&&+\A_y^2y^2(f)+ D_\psi^2 \psi^2(f)\\
&&+\A_yD_\psi\frac{4k_BT_0}{\w^2}\Re(Y_{y\psi}+Y_{\psi y})\\
&&+\A_z^2z^2(f) + D_\phi^2 \phi^2(f)
\end{eqnarray*}
so it is a weighted sum in
quadrature of the thermal noise of different degrees of freedom, plus
some cross-terms. These terms may be negative, so it is possible to
choose an optimal set of parameters to minimize the sensed motion, as
shown in \cite{Yuri}.
The weighting factors and distances are
\begin{eqnarray*}
\A_x&=&\cos(\gamma_y)\cos(\gamma_z)\\
\A_y&=&\sin(\gamma_y)\cos(\gamma_z)\\
\A_z&=&\sin(\gamma_z)\\
D_\theta&=&R\sin(\gamma_z)+dz\cos(\gamma_y)\cos(\gamma_z)\\
D_\phi&=&\cos(\gamma_z)(d_y\cos(\gamma_y)-R\sin(\gamma_y))\\
D_\psi&=&d_y\sin(\gamma_z)+d_z\sin(\gamma_y)\cos(\gamma_z)
\end{eqnarray*}

Typical distances $d_y,d_z$ are few mm at most, and typical angles
$\gamma_y$are in the order of microradians, since an angle of such
magnitude produces displacements of the order of millimeters at the
beam at the other end of the arm, 4km away. However, the angle
$\gamma_z$ cannot be less than half the arm length divided by the
curvature of Earth, or $\gamma_z\geq 3\times 10^{-4}$. We show in
Fig\ref{FinalPlot} the thermal noise sensed by a beam with
$d_y=d_z=5$mm , $\gamma_z=3\times 10^{-4}$ and $\gamma_y=1\mu$rad.

\section{Results and conclusions}

We have shown several general results, and then calculated predicted
thermal motions for LIGO suspensions. First, we showed that if there is
a single oscillator with two sources of potential energy, one with a
spring constant $K_g$ and a dominant real part, and the other with a
complex spring constant $K_e$, with a dominant loss factor, then the
dilution factor $K_e/K_g$ gives us the ratio between the oscillator's
quality factor (determining its thermal noise spectral density) and
the loss factor of $K_e$. However, the elastic loss
$\phi_e=\Im{K_e}/\Re{K_e}$ might itself have a small dilution factor
with respect to the loss factor of the Young modulus, for example if
$K_e\sim\sqrt{E}$, then there is a dilution factor of 1/2.

We also show that when two or more degrees of freedom are coupled,
the measurable quality factors at resonance may not be as useful to
predict the ``effective'' quality factor used in the thermal noise
spectral density, unless the eigenfrequencies are far from each other.

We showed that by using approximations of order $e^{-L/\D}$, we can
easily obtain wire shapes and equations of motion for the 6 degrees of
freedom of the mirror, as a function of applied oscillating forces and
torques. We think that this method will be most useful when applied to
multiple pendulum systems such as those used in GEO600 and planned for
advanced LIGO detectors \cite{detectors,LIGO}. However, even in the simple
pendulum case this approach allows us to calculate the gravitational
and elastic potential energy as a linear energy density along the
wire, and the total energy. We showed that, for an applied horizontal
force, the gravitational potential energy ($(1/2)\int Tw'^2 ds$) is
homogeneously distributed along the wires, while the elastic potential
energy ($(1/2)\int EI (w'')^2 ds$) is concentrated at the top and
bottom, but in different proportions depending on the frequency of the
applied force (Fig. \ref{CumulativeEnergies}. We also calculate the
ratio of total elastic potential energy to gravitational energy using
the solutions for the wire shape, and show that this function of
frequency corresponds to the dilution factor for the eigenmode loss
factors {\em as well} as for the effective quality factor that allows
us to calculate the thermal noise at gravitational wave frequencies
(see Figs.\ref{PtMass},\ref{EnergyRatios},\ref{YawDilutionFactor}).

Applying our calculation of wire shapes and equations of motion that
include elasticity to LIGO suspensions, we show in Figs \ref{disptn}
and \ref{angtn} the resulting spectral densities of displacement and
angular degrees of freedom of the mirror. More importantly, we show in
Fig\ref{FinalPlot} the resulting contribution of pendulum thermal
noise to the LIGO sensitivity curve, assuming small misalignments in
the sensing laser beam (5mm away from center of mass, 1$\mu$rad away
from horizontal). As expected, the displacement degree of freedom is
the one that dominates the contribution, but pitch noise contributes
significantly (29\% at 100 Hz) if the beam is 5mm above center. But
this is not added in quadrature to the horizontal noise, which makes
81\%: the coupled displacement-pitch thermal motion makes up 99\% of
the total thermal noise. A misplacement {\em below} the center of mass
will {\em reduce} the observed thermal noise, as first noted in
\cite{Yuri}. The contribution of yaw thermal noise (11\% at 100Hz) is
smaller than that of pitch, but very comparable. The contribution of
vertical noise due to the 4km length of the interferometer is 8\% at
100 Hz. The side and roll motions are coupled, but the roll
contribution dominates (due to the large angle $\gamma_z$ and the
assumed $d_z=5$mm) and is 0.7\%, much smaller than the contributions
of pitch, yaw and vertical degrees of freedom. When added in
quadrature, the total thermal noise is 23\% higher than the
contribution of just the horizontal thermal noise. However, if the
vertical misplacement of the beam is 5mm below the center of mass,
instead of above, the total contribution of thermal noise is 89\% of
the horizontal thermal noise of the center mass.

%%%%%%%%%%%%%
\section{Acknowledgments}
Much of this work was motivated by many discussions held with Jim
Hough, Sheila Rowan and Peter Saulson, and I am very glad to thank
their insights. I want to especially thank P. Saulson for carefully
reading the original manuscript and making important suggestions. I
also want to thank P. Fritschel and Mike Zucker, who first asked about
thermal noise of angular modes in pendulums. This work was supported
by NSF grants 9870032 and 9973783, and by The Pennsylvania State
University.

%%%%%%%%%%%%%%%%%%%%%%%
%%%%%%%%%%%%%%%%%%%%%%%
%%%%%%%%%%%%%%%%%%%%%%%

}\end{multicols}

%%%%%%%%%%%%%%%%%%%%%%%%%%%%%%%%%%%%%%%%%%%%%%%%%%%%%%%%%%%%

\newpage

\begin{figure}
\centerline{\hbox{\psfig{figure=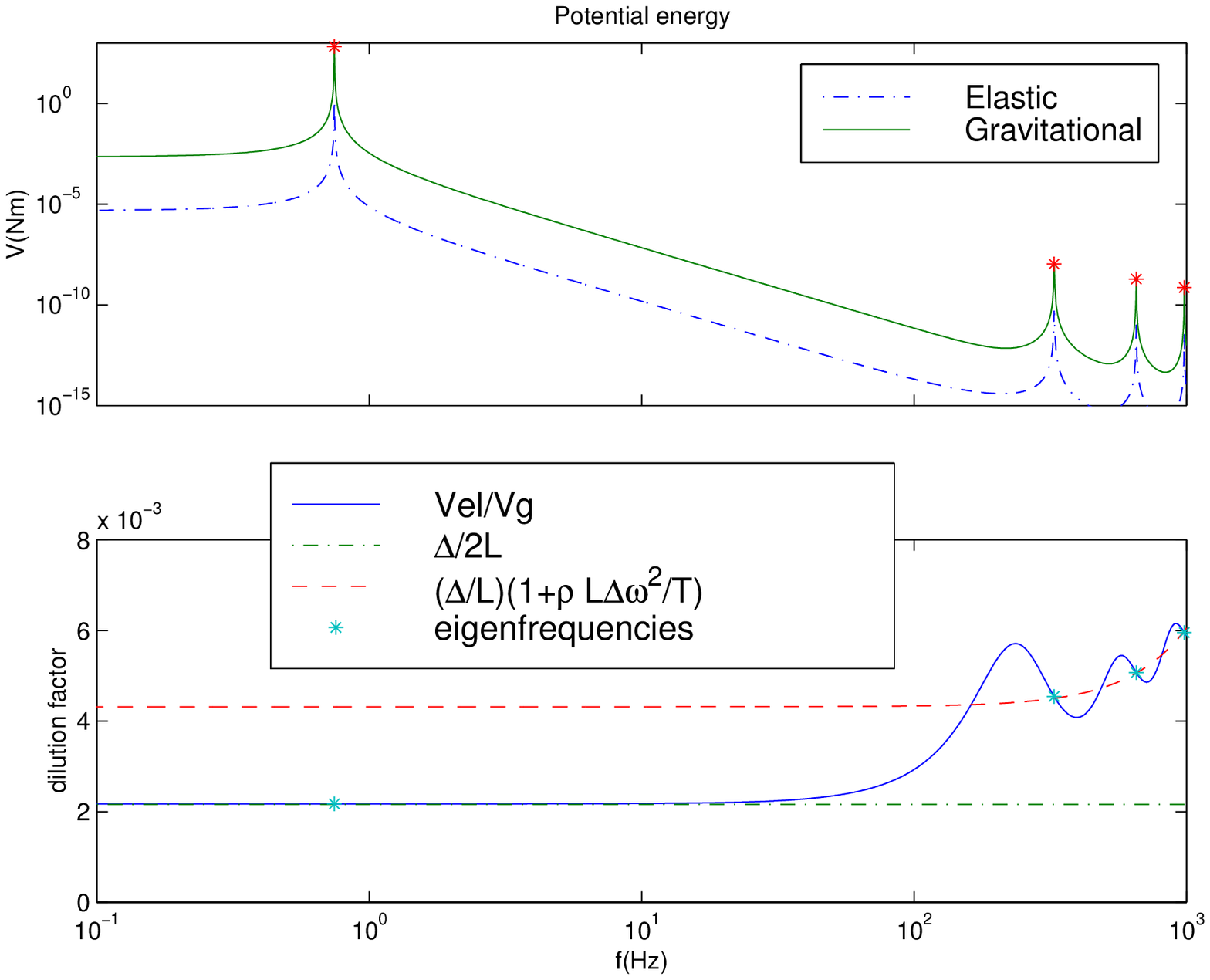,height=5in}}}
\caption{Top figure: Gravitational and potential energies for a
suspended point mass as a function of frequency $f$, when excited by a
sinusoidal force wiht frequency $f$. Bottom figure: ratio of energies,
considered as a ``dilution factor''. The stars indicate the factor
$\phi/Q_i$, where $\phi$ is the wire loss factor and $Q_i$ is the
eigenmode quality factor.}
\label{PtMass}
\end{figure}

\begin{figure}
\centerline{\hbox{\psfig{figure=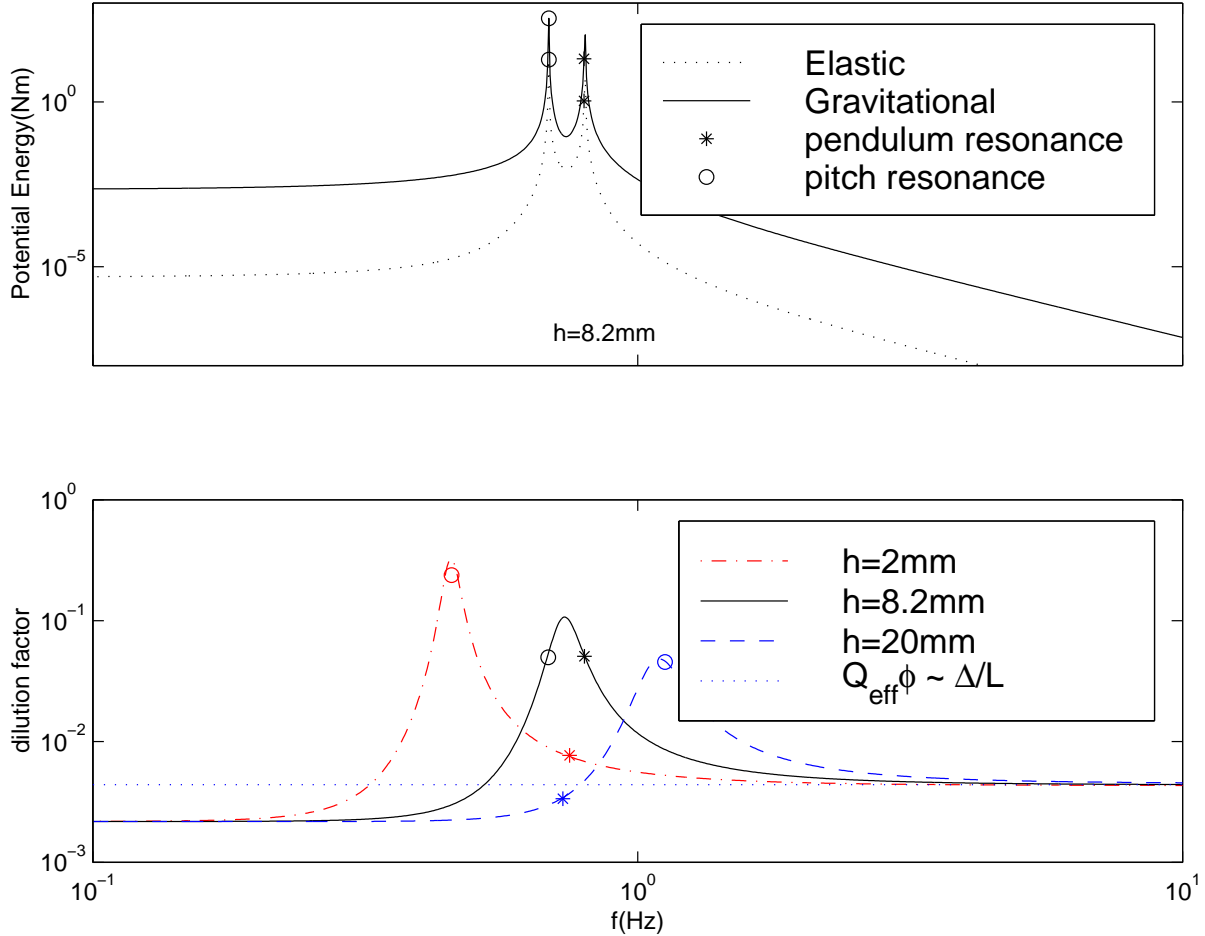,height=5in}}}
\caption{At the top, we plot the elastic and gravitational potential
energies for a pitch distance h=8.2mm. In the bottom figure, we plot
the ratio of elastic and gravitational energy for three different
values of the pitch distance. The stars represent the dilution factor
of the pendulum mode, at the pendulum frequency; the circles are the
dilution factors of the pitch mode, at the pitch eigenfrequency; and
the solid horizontal line is the dilution factor of the effective
quality factor at 50 Hz.}
\label{EnergyRatios}
\end{figure}

\begin{figure}
\centerline{\hbox{\psfig{figure=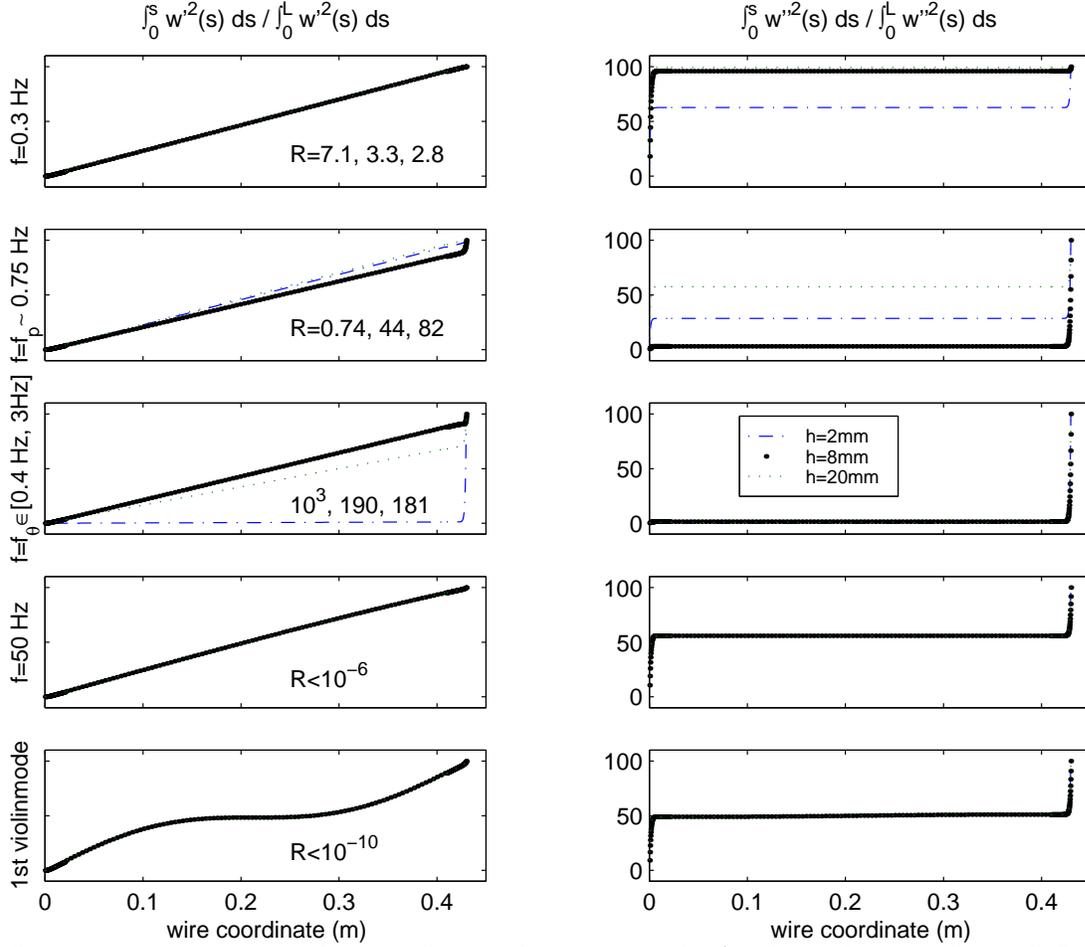,height=5in}}}
\caption{The figure shows cumulative potential energies along the wire
(top is $s=0$, bottom is $s=0.45$m), when the pendulum is excited by a
force a the center of mass of frequency $f$. The energies are shown
for three different pitch distances in each graph: h=2mm (dashed),
h=8mm (thick points) and h=20mm (dotted). The energies are also
plotted for five different frequencies: $f=0.3$ Hz, below all
eigenmodes; $f=f_p$ orpendulum frequency, approximately 0.75 Hz;
$f=f_\T=\sqrt{Th/J}/(2\pi)=(0.46,0.69,1.12)$Hz for $h=(2,8,20)$mm
respectively; f=50Hz, in the gravitational wave band; and $f=319$Hz,
the first violin mode frequency.}
\label{CumulativeEnergies}
\end{figure}

\begin{figure}
\centerline{\hbox{\psfig{figure=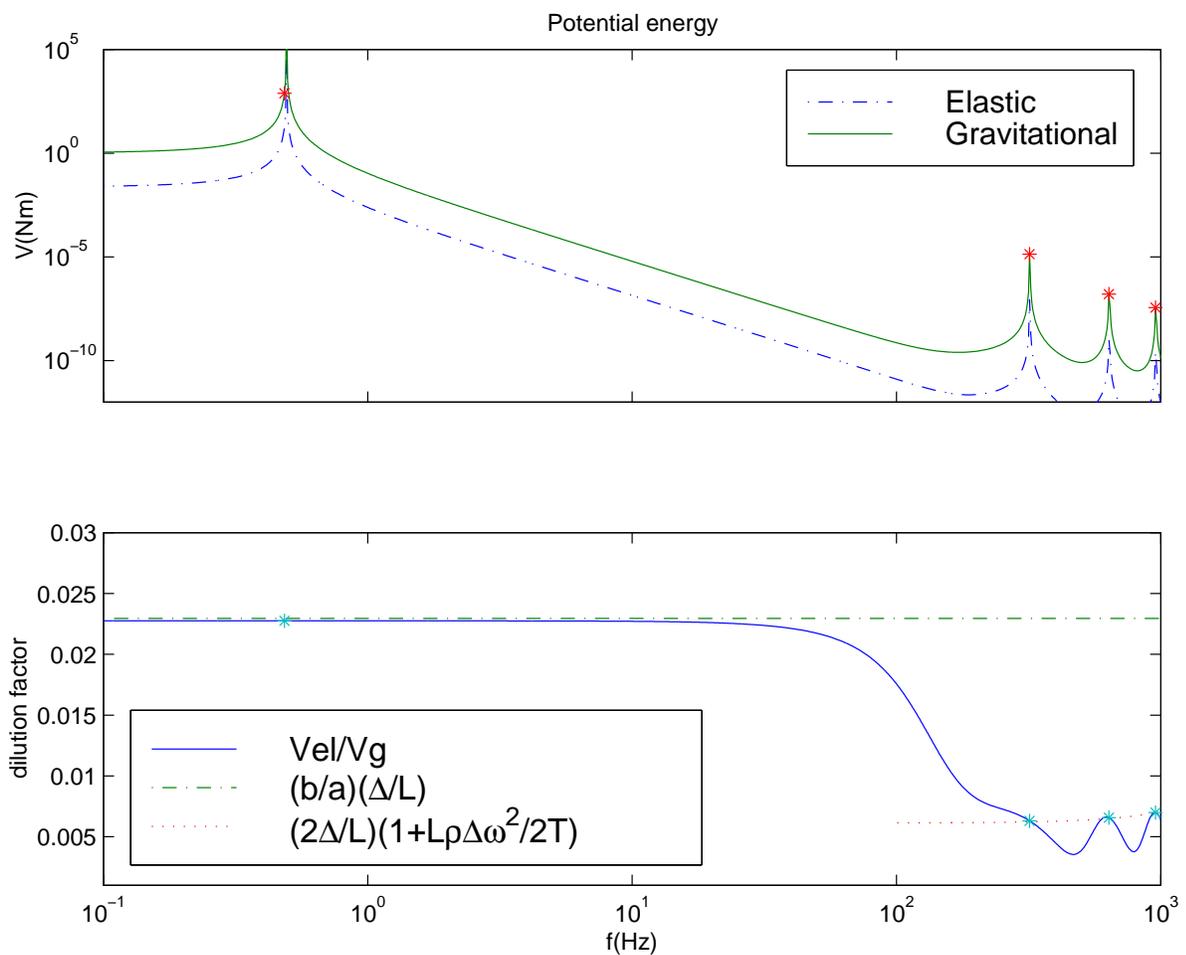,height=5in}}}
\caption{Potential energies and dilution factor for LIGO yaw degree of freedom}
\label{YawDilutionFactor}
\end{figure}

\begin{figure}
\centerline{\hbox{\psfig{figure=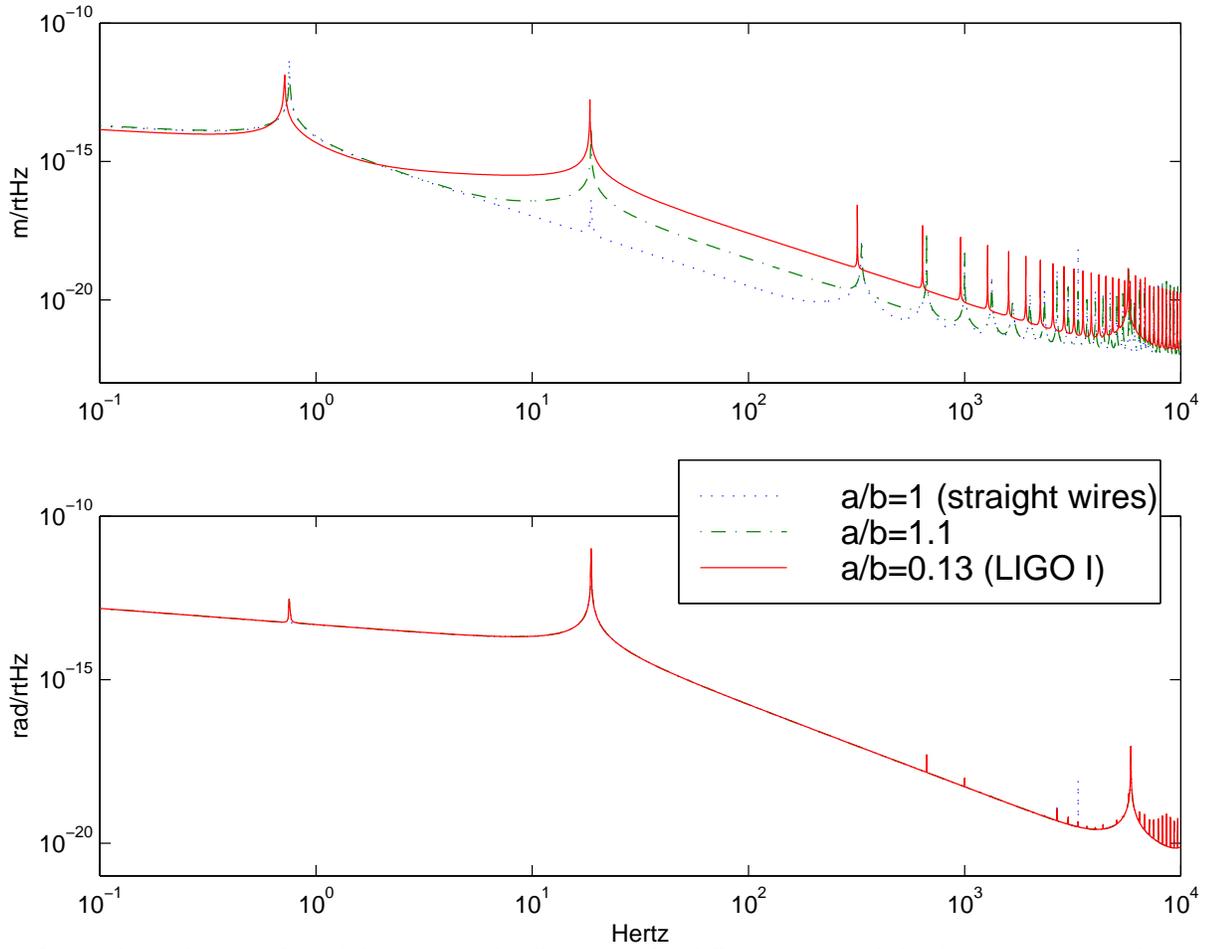,height=5in}}}
\caption{Thermal noise of pendulum side and roll motion, for different
values of the distance between the wires at the top. The wires are
tilted with respect to the vertical direction at an angle
$\sin\A=(b-a)/L$, where $L$ is the length of the wires. }
\label{sideroll}
\end{figure}

\begin{figure}
\centerline{\hbox{\psfig{figure=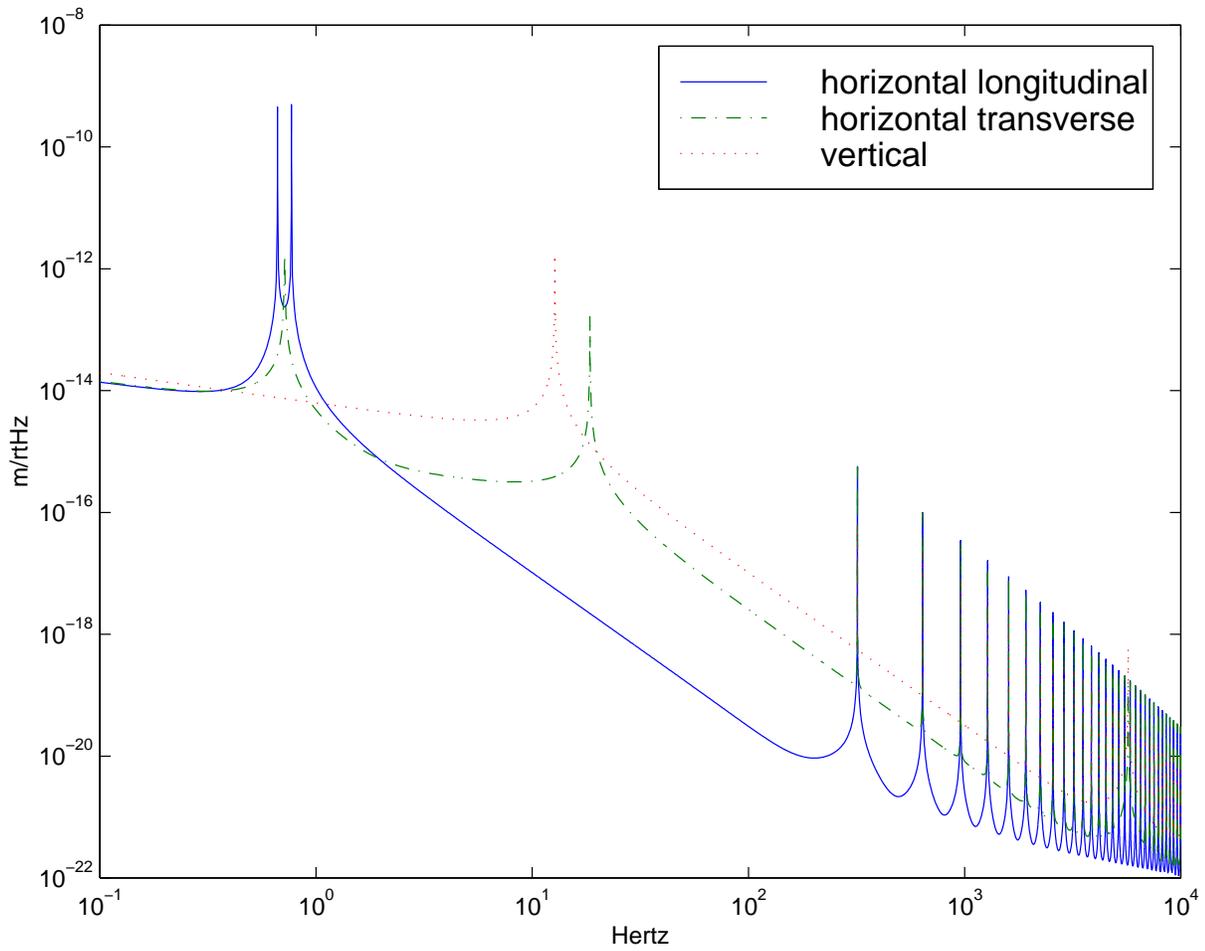,height=5in}}}
\caption{Thermal noise of the three mirror's displacement degrees of freedom.}
\label{disptn}
%Produced with DispAngtn.m 
\end{figure}

\begin{figure}
\centerline{\hbox{\psfig{figure=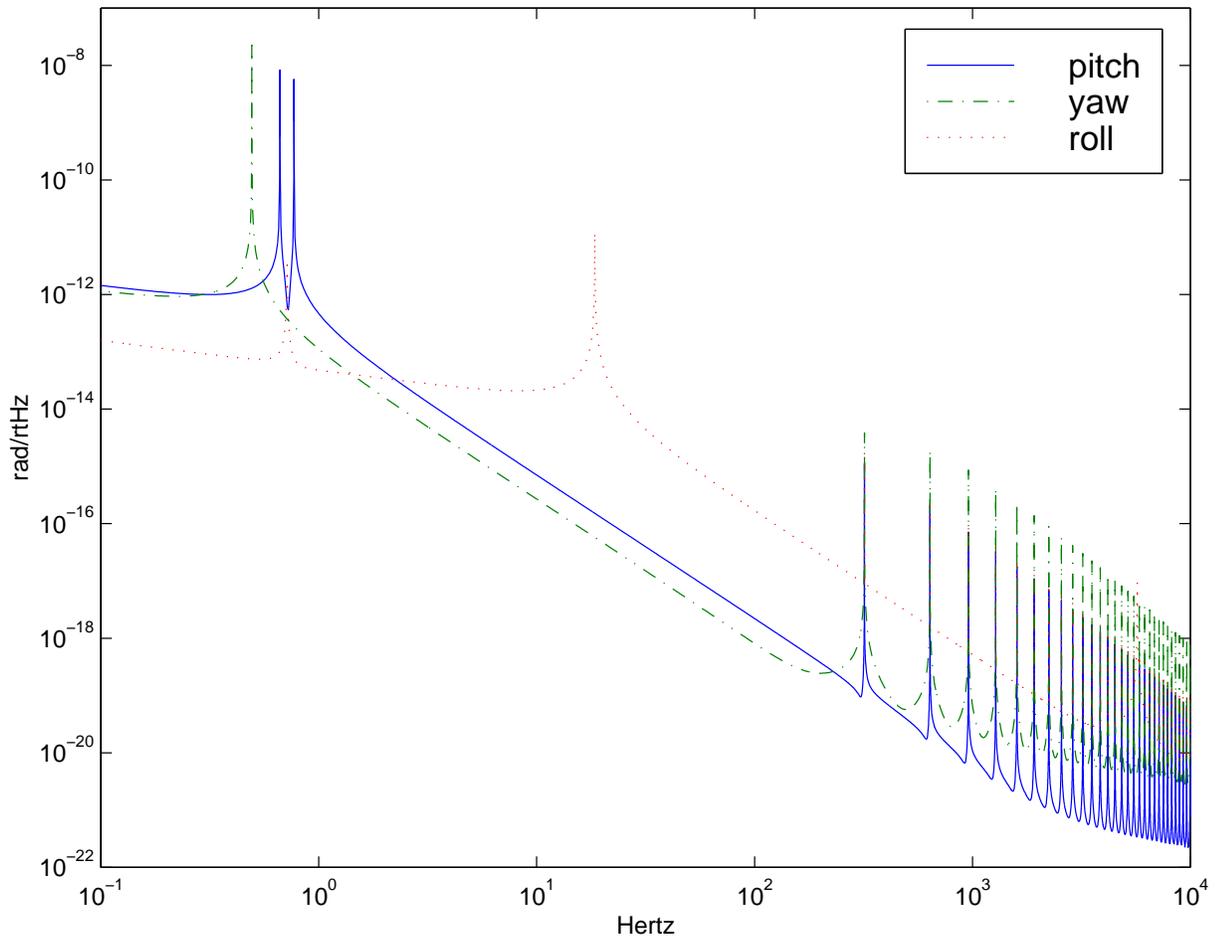,height=5in}}}
\caption{Thermal noise of the three mirror's angular degrees of freedom.}
\label{angtn}
%Produced with DispAngtn.m 
\end{figure}

\begin{figure}
\centerline{\hbox{\psfig{figure=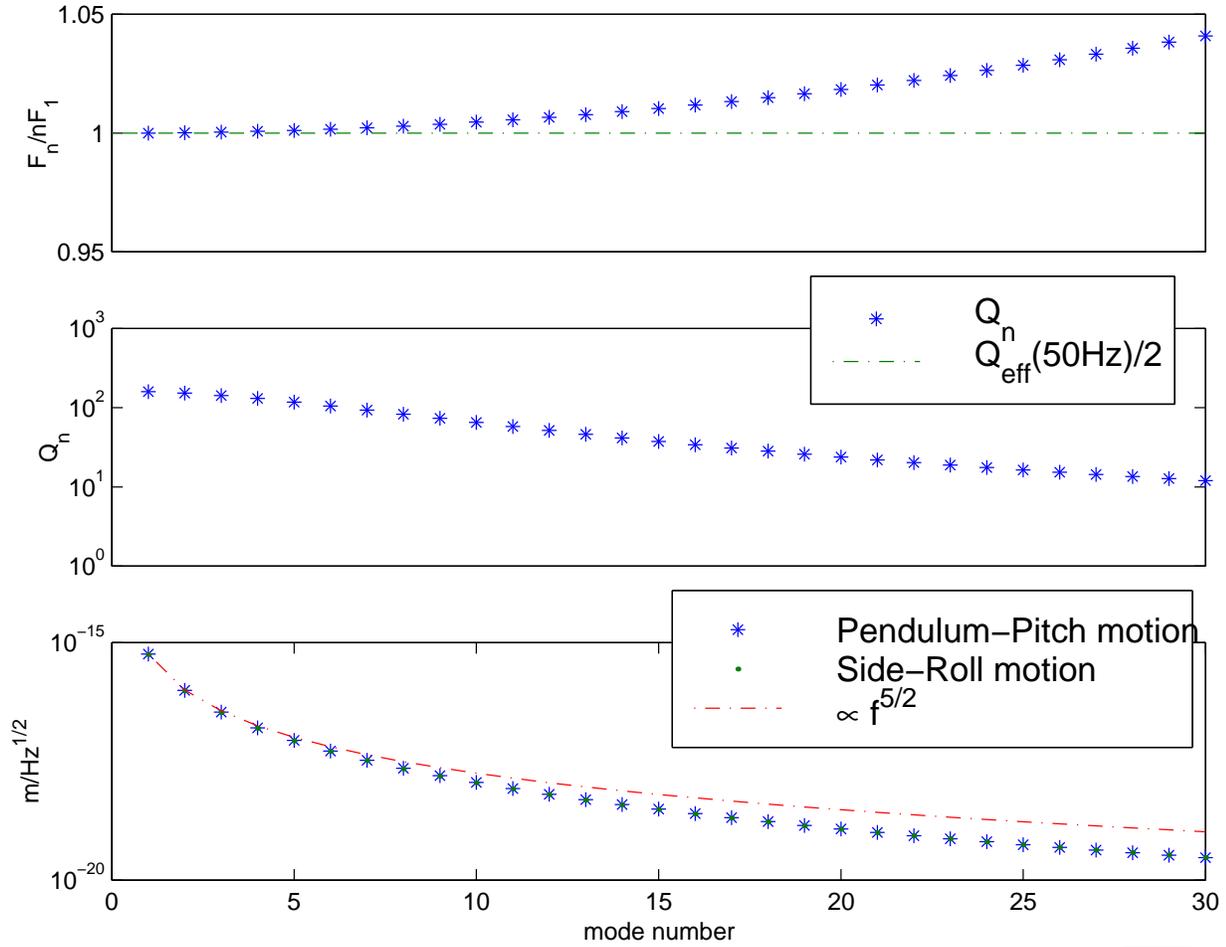,height=5in}}}
\caption{In the top figure, we show the frequencies of the resonant
violin modes, scaled by $f_n=n\sqrt{T/rho}/2L$, for longitudinal and
transverse modes. In the middle figure, we show the corresponding
quality factors. The solid line is $Q_eff/2\sim L/2\D$. In the bottom
figure, we show the amplitude of longitudinal and transverse thermal
noise at the violin modes. The solid line shows a $1/f^{5/2}$ fall off
from the first peak.}
\label{ViolinModes}
\end{figure}

\begin{figure}
\centerline{\hbox{\psfig{figure=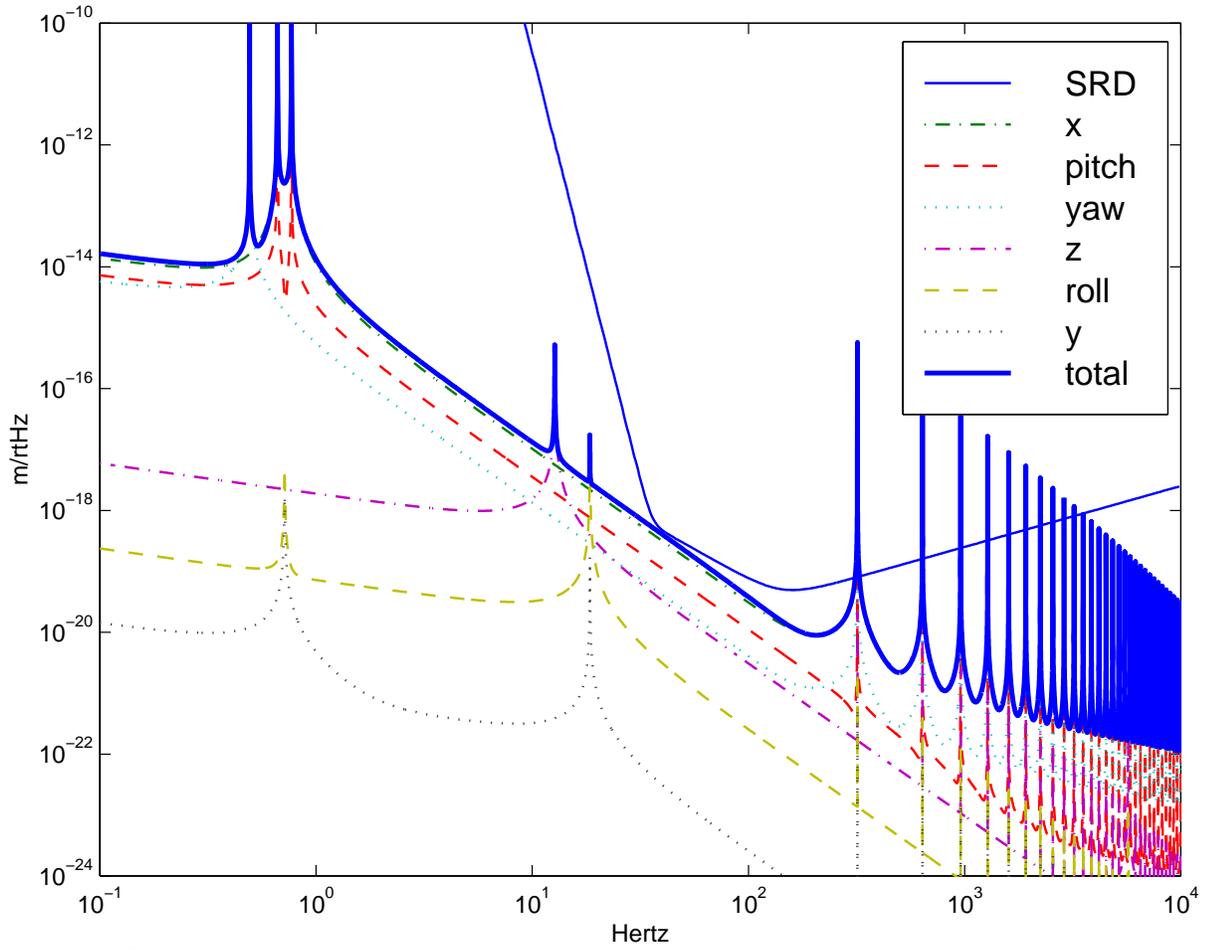,height=5in}}}
\caption{Thermal noise sensed by a laser beam 5mm away from center,
and 1$\mu$rad away from horizontal. The ``SRD'' curve is the expected
sensitivity of LIGO I. The individual degrees of freedom are plotted
in their order of contribution at 100 Hz.}
\label{FinalPlot}
\end{figure}

\end{document}